\documentclass[acmtog]{acmart}
\acmSubmissionID{204}

\usepackage{booktabs} % For formal tables

\usepackage{graphicx}
\usepackage{grffile}
\usepackage{capt-of}
\usepackage{xcolor}
\usepackage{amsmath,amsfonts,dsfont,pifont,bm,bbm,mathrsfs,mathtools,nicefrac} % amssymb causes compile error
\usepackage{algpseudocode,listings} % algorithm causes compile error
\usepackage{booktabs,multirow,adjustbox,diagbox,threeparttable}
\usepackage{subcaption}
\usepackage{multirow}
\usepackage{array}
\newcolumntype{L}[1]{>{\raggedright\let\newline\\\arraybackslash\hspace{0pt}}m{#1}}
\newcolumntype{C}[1]{>{\centering\let\newline\\\arraybackslash\hspace{0pt}}m{#1}}
\newcolumntype{R}[1]{>{\raggedleft\let\newline\\\arraybackslash\hspace{0pt}}m{#1}}

\usepackage{enumitem}
\setlist{leftmargin=18pt}

\acmJournal{TOG}

\usepackage{fp}
\usepackage{expl3}[2012-07-08]
\ExplSyntaxOn
\ExplSyntaxOff

\usepackage{amsmath,amsfonts,bm}

\newcommand{\ignorethis}[1]{}

\def\eqref#1{equation~\ref{#1}}

\def\1{\bm{1}}

\def\ie{{\textit{i.e. }}}

\def\rva{{\mathbf{a}}}

\def\rvd{{\mathbf{d}}}
\def\rve{{\mathbf{e}}}

\def\rvg{{\mathbf{g}}}

\def\rvl{{\mathbf{l}}}
\def\rvm{{\mathbf{m}}}
\def\rvn{{\mathbf{n}}}
\def\rvo{{\mathbf{o}}}
\def\rvp{{\mathbf{p}}}
\def\rvq{{\mathbf{q}}}
\def\rvr{{\mathbf{r}}}
\def\rvs{{\mathbf{s}}}

\def\rvv{{\mathbf{v}}}

\def\rvx{{\mathbf{x}}}

\def\rvz{{\mathbf{z}}}

\DeclareMathAlphabet{\mathsfit}{\encodingdefault}{\sfdefault}{m}{sl}
\SetMathAlphabet{\mathsfit}{bold}{\encodingdefault}{\sfdefault}{bx}{n}

\def\sR{{\mathbb{R}}}

\DeclareMathOperator{\sign}{sign}

\newcommand{\ignore}[1]{}

\newcommand*\diff{\mathop{}\!\mathrm{d}} % dx in integeral

\setcopyright{rightsretained}
\acmJournal{TOG}
\acmYear{2024} \acmVolume{43} \acmNumber{6} \acmArticle{254} \acmMonth{12}\acmDOI{10.1145/3687768}

\begin{document}

% Title portion
\title{GroomCap: High-Fidelity Prior-Free Hair Capture}

\author{Yuxiao Zhou}
\affiliation{%
  \institution{ETH Zurich}
  \country{Switzerland}}
\email{yuxiao.zhou@inf.ethz.ch}

\author{Menglei Chai}
\affiliation{%
  \institution{Google Inc.}
  \country{United States of America}}
\email{mengleichai@google.com}

\author{Daoye Wang}
\affiliation{%
  \institution{Google Inc.}
  \country{Switzerland}}
\email{daoye@google.com}

\author{Sebastian Winberg}
\affiliation{%
  \institution{Google Inc.}
  \country{Switzerland}}
\email{winbergs@google.com}

\author{Erroll Wood}
\affiliation{%
  \institution{Google Inc.}
  \country{United Kindom}}
\email{errollw@google.com}

\author{Kripasindhu Sarkar}
\affiliation{%
  \institution{Google Inc.}
  \country{Switzerland}}
\email{krsarkar@google.com}

\author{Markus Gross}
\affiliation{%
  \institution{ETH Zurich}
  \country{Switzerland}}
\email{grossm@inf.ethz.ch}

\author{Thabo Beeler}
\affiliation{%
  \institution{Google Inc.}
  \country{Switzerland}}
\email{tbeeler@google.com}

\begin{abstract}

Despite recent advances in multi-view hair reconstruction, achieving strand-level precision remains a significant challenge due to inherent limitations in existing capture pipelines.
We introduce \textit{GroomCap}, a novel multi-view hair capture method that reconstructs faithful and high-fidelity hair geometry without relying on external data priors.
To address the limitations of conventional reconstruction algorithms, we propose a neural implicit representation for hair volume that encodes high-resolution 3D orientation and occupancy from input views.
This implicit hair volume is trained with a new volumetric 3D orientation rendering algorithm, coupled with 2D orientation distribution supervision, to effectively prevent the loss of structural information caused by undesired orientation blending.
We further propose a Gaussian-based hair optimization strategy to refine the traced hair strands with a novel chained Gaussian representation, utilizing direct photometric supervision from images.
Our results demonstrate that \textit{GroomCap} is able to capture high-quality hair geometries that are not only more precise and detailed than existing methods but also versatile enough for a range of applications.

\end{abstract}

\begin{CCSXML}
<ccs2012>
<concept>
<concept_id>10010147.10010371.10010396.10010399</concept_id>
<concept_desc>Computing methodologies~Parametric curve and surface models</concept_desc>
<concept_significance>500</concept_significance>
</concept>
</ccs2012>
\end{CCSXML}
\ccsdesc[500]{Computing methodologies~Parametric curve and surface models}

\keywords{Strand-level hair modeling, multi-view reconstruction}

\begin{teaserfigure}
\includegraphics[width=\linewidth]{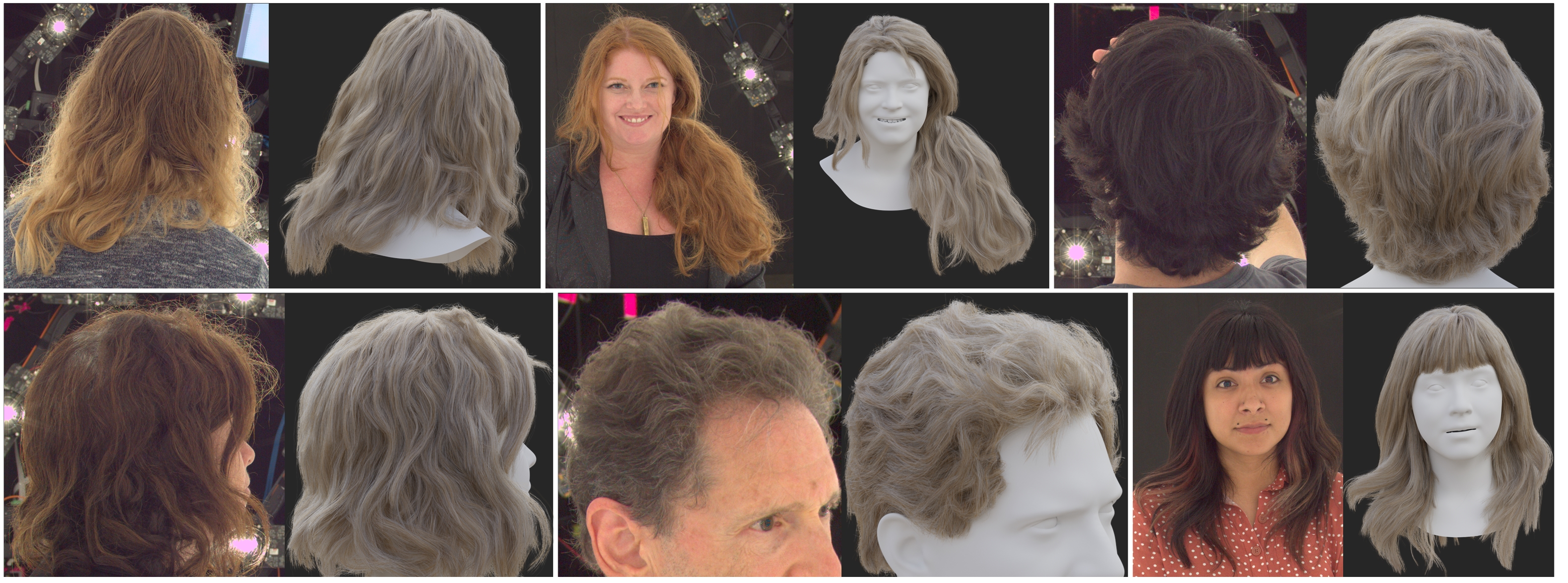}
\caption{\textbf{\textit{GroomCap} reconstructs high-fidelity hair geometry across a diverse array of hairstyles.} For each result, we show one input view on the left, alongside a rendering of the reconstructed hair model from the same view on the right. A pre-defined material is used to better visualize geometric details.}
\label{fig:teaser}
\end{teaserfigure}

\maketitle

\section{Introduction}

Hairstyles are not merely aesthetic decorations; they serve as a profound expression of individual and cultural identity that shapes our perception of others.
In the digital realm, realistic hair plays a crucial role in virtual reality, gaming, and for digital doubles, where visual authenticity is paramount.
Despite ongoing efforts in human digitization, accurate reconstruction of strand-level hair geometry from images remains a formidable and unique challenge. This complexity arises from the detailed, layered, and intertwined structures of hair, alongside its natural variability in style, texture, and material, all of which significantly complicate the capture process.

Over the past decades, multi-view hair reconstruction has seen encouraging progress. Well before the advent of neural networks, traditional capture pipelines typically reconstruct the visible exterior of hair and compute a 3D orientation field using 2D estimates from each view, followed by completing the 3D structure of the entire hair volume to generate the final strands.
Despite the promising results, the faithfulness and fidelity of the reconstructed hair are largely compromised by key limitations of these pipelines: reliance on explicit point clouds and orientation projections may overlook critical details in regions of uncertainty and overlapping strands; simple volume completion like diffusion or ribbon conversion often produces overly smoothed hair and may only work effectively on distinct wisp structures; and the final strand extraction further deviates the result from the input.

To address these challenges, recent methods have sought to leverage data-driven solutions, incorporating priors learned from synthetic data into hair capture pipelines. While prior-based reconstruction has seen notable success in other human components, such as faces, hands, and bodies, its application to hair remains challenging for two primary factors. First, the scarcity of ground-truth hair data necessitates large-scale synthetic assets, which are not only costly to collect at scale but also suffer from an inevitable gap to real-world examples. Second, due to the extreme diversity and variability of hair, even the most extensive hair libraries cannot adequately cover the precise details of specific subjects. Consequently, despite more visually pleasing results, these prior-based approaches often yield highly regularized and flattened geometries that struggle to capture structural details outside the training set.

Data priors are not a cure-all. Simply plugging prior models into capture pipelines, without addressing the underlying algorithmic limitations, still results in challenges that hinder us from achieving high-quality hair geometry capture. In this work, we take a more fundamental look at the problem and push the boundaries of high-quality hair capture \textit{without relying on any data priors}.

By examining existing capture pipelines, we identify several key issues:
1) Representing hair as discrete exterior surfaces or explicit volumes often leads to a significant loss of spatial information, such as rich structural details and natural variations in occupancy.
2) 3D hair structure is inferred from image-based orientation estimation, where each pixel is formed from superposing numerous strands with dramatically different directions. However, the common practice of aggregating them into a single orientation angle discards crucial structural information necessary for accurate recovery.
3) Even with initially accurate projected 3D structures, the quality of the hair geometry tends to degrade after volume completion and strand tracing, resulting in poor spatial distribution, missing local details, inconsistent boundaries, and unnatural curvatures.

In this work, we introduce \textit{GroomCap}, a novel multi-view hair capture pipeline aimed at reconstructing high-fidelity and strand-level accurate hair geometry without external data priors.
Our method incorporates several major technical innovations leading to unprecedented performance, versatility, and robustness.
Firstly, we propose a neural implicit representation for volumetric hair, encoding 3D orientation and occupancy from input views. Compared to exterior surfaces or explicit volumes, our implicit hair volume enjoys greater accuracy, expressiveness, and memory-efficiency.
Secondly, we train our hair volume model to effectively capture the complete hair structure. To achieve this, we develop a new volumetric 3D orientation rendering algorithm, where orientation integration is performed along each ray, maintaining all overlapping hair structures without blending. Correspondingly, we revisit 2D orientation estimation to estimate a per-pixel \textit{orientation distribution} as the training supervision, rather than a single orientation angle.
Finally, we introduce Gaussian-based hair optimization, applied to initial hair strands traced from the volume, to improve their faithfulness and fidelity through direct photometric supervision from input images. The key ingredient is a new chained hair Gaussian representation, featuring carefully tailored geometry and appearance parameters, along with a dynamic splitting and pruning mechanism.

Altogether, GroomCap effectively captures accurate and high-fidelity dense hair models for a diverse range of hairstyles, using the same pipeline and parameters for all of them. The resulting strand geometries are consistently natural and guaranteed to be scalp-rooted, thus supporting various editing applications, including re-rendering, physics-based animation, and interactive grooming.

\section{Related Work}

\subsection{Hair Capture without Data Prior}

Multi-view hair capture is of great interest to both research and industry communities, in attempts to digitize 3D hair without involving laborious artistic authoring.
Early efforts \cite{paris04capture,wei05modeling} construct visual hulls to constrain the hair volume and estimate 3D hair orientations that are consistent across views. \cite{paris08hair} proposes a system that is capable of reconstructing exterior strand positions and growing strands within diffused orientation volumes. \cite{luo12multi,luo13wide} introduce multi-view stereo methods for reconstructing detailed hair surfaces using 2D orientation fields. Following that, \cite{luo13structure} proposes a structure-aware hair capture method that incorporates structural priors to predict ribbon connectivity for capturing hair wisp structures.
Focusing on capturing sparse outer strand segments instead of complete hairstyles, \cite{jakob09capturing} detects accurate hair fibers with shallow depth of field captures. \cite{nam19strand} introduces line-based PatchMatch multi-view stereo for hair, which is further improved by \cite{sun21human} to support hair inverse rendering.

Besides RGB images, other imaging modalities have also been investigated, such as RGB-D \cite{zhang18modeling} and thermal imaging \cite{herrera12lighting}. Recently, \cite{shen2023CT2Hair} achieves high-quality hair reconstructions by leveraging computed tomography (CT) scans to obtain inner strand structures. However, it is unsuitable for use on live human subjects due to the large exposure of X-rays.

Instead of strand-based geometry, some recent methods \cite{alexandru22neural,wang22hvh,wang23neuwigs} implicitly reconstruct hair in volumetric representations. Despite achieving great visual quality, their primary focus is on hair image synthesis for novel views or motions, rather than capturing precise strand geometry itself.

\subsection{Model-Based Hair Capture}

Data priors start to gain traction in single-view \cite{chai12single,chai13dynamic,chai15high,chai16autohair,hu15single} or sparse-view hair modeling \cite{zhang17data} to address the inherent challenges of these highly ill-posed problems.
Early works \cite{hu15single,chai16autohair} focus on matching and retrieving the closest dataset items to the input views as the foundation for further fitting. More recently, neural-based approaches become dominant the field, offering improved accuracy and robustness.
For example, \cite{zhou18hairnet} trains a convolutional neural network to infer geometry encoding from input images.
\cite{saito183d} develops a volumetric variational autoencoder for generating hair conditioned on the input image.
\cite{yang19dynamic} infer 3D shape and motion from monocular videos for dynamic hair capture.
\cite{wu22neuralhdhair} trains a voxel-aligned implicit function to infer 3D volumetric information from the image.
\cite{zheng23hairstep} proposes inferring 2D depth and orientation maps before predicting 3D geometry.
\cite{kuang22deepmvshair} extends these techniques to handle sparse view inputs.

In the context of dense multi-view hair capture, early efforts adopt data priors to enhance reconstruction robustness with physics-based strand priors \cite{hu14robust} or address the specific challenge of braid reconstruction with pre-defined prior models on braid patterns \cite{hu14capturing}.
More recently, researchers have begun to integrate data priors more deeply into hair capture pipelines. 
\cite{sklyarova23neural} employs a surface-based representation for the coarse shape of the hair volume and reconstructs hair strands as a geometry texture in a prior-guided manner.
\cite{wu24monohair} combines data priors for predicting interior structure with conventional patch-based multi-view optimization for exterior reconstruction, achieving state-of-the-art results.

\subsection{Preliminaries}

\paragraph{Neural implicit fields}
Neural implicit fields leverage neural networks to represent 3D scenes.
In the groundbreaking work of NeRF \cite{mildenhall2021nerf}, a multi-layer perceptron (MLP) model is adopted to predict volume density and radiance for any 3D point.
Using this MLP, each pixel can be rendered through alpha-blending the radiance of sample points along the ray, akin to classical volume rendering techniques. This allows for the synthesis of arbitrary views by rendering all rays originating from a virtual camera.
The model itself is trained on multi-view images, adhering to the synthesis-and-comparison paradigm.
Our method builds upon the original NeRF model, extending its capabilities to 3D hair structure reasoning.

\paragraph{3D Gaussian splatting}
3D Gaussian splatting (3DGS) \cite{kerbl20233d} is a recent technique that represents a scene using anisotropic 3D Gaussians. Each Gaussian is defined by a set of parameters, including position, covariance matrix, opacity, and color.
These Gaussians can be rendered differentiably using the splatting method and are trained with supervision from color images.
While 3DGS is effective at reproducing visual appearance, the underlying geometry formed by these Guassians has been less explored.
In this paper, we render hair strands as chained Gaussians, employing pixel-wise supervision. Additionally, we propose a novel fomulation that enforces geometric constraints during the optimization.

\section{Method Overview}

\begin{figure}[t]
  \includegraphics[width=\linewidth]{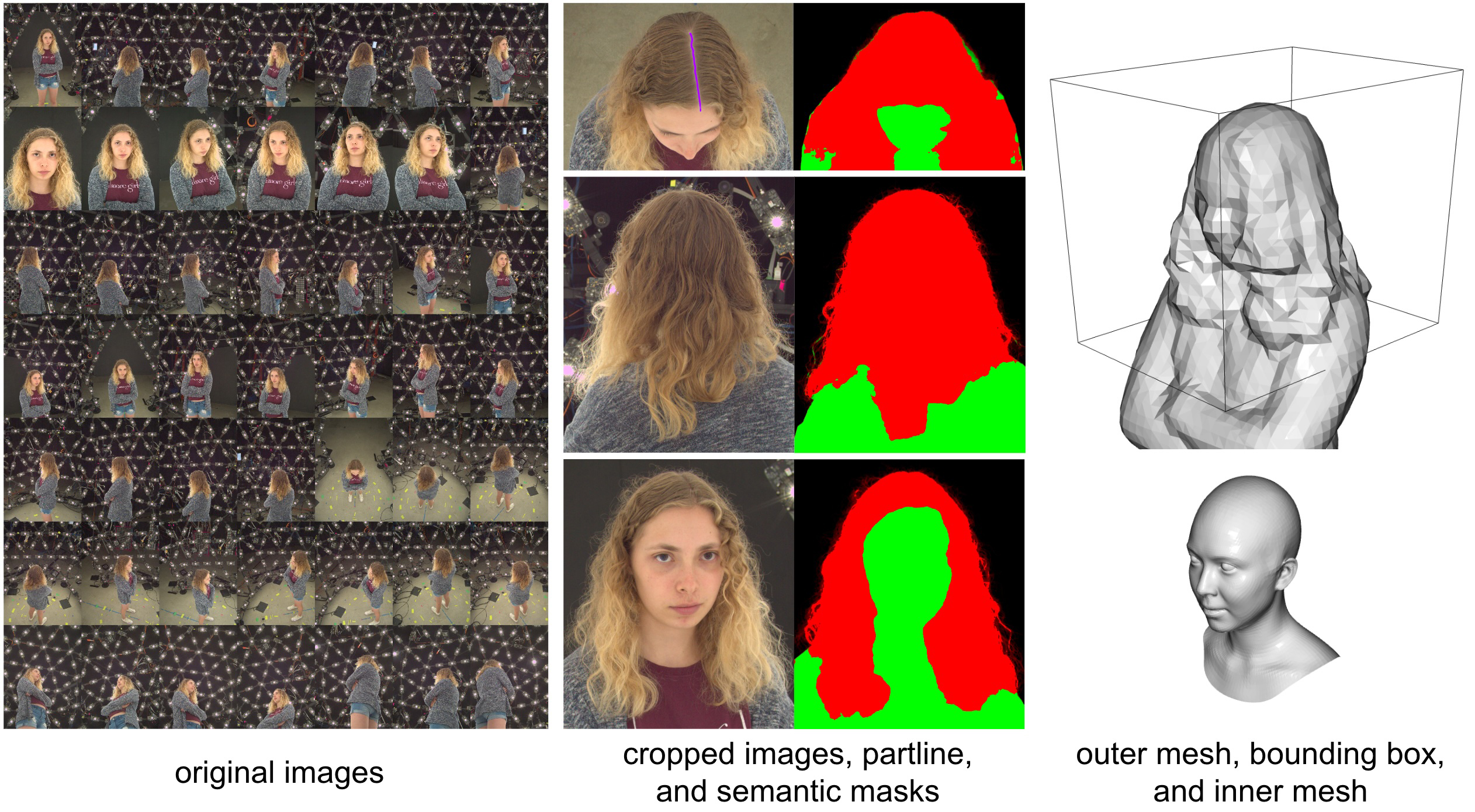}
  \caption{\textbf{The input to our pipeline} includes calibrated multi-view images (left), semantic segmentations of hair and foreground (middle), reconstructed inner and outer meshes with the hair bounding box (right), and optional hair partline annotation on one image (middle column, first row).}
  \label{fig:inputs}
\end{figure}

\subsection{Our Pipeline}
\label{subsec:overview:pipeline}
Our method contains three stages.
In the first stage (\autoref{sec:volume}), we establish an \textit{implicit hair volume} that encodes both the spatial occupancy and orientation of the target hairs from multi-view image captures (Sec.~\ref{subsec:overview:data}).
In the second stage (\autoref{sec:tracing}), we grow \textit{initial hair strands} within the hair volume based on simple heuristics.
In the last stage (\autoref{sec:3dgs}), starting from these initial strands, we optimize the \textit{final hair geometry} with respect to multi-view images utilizing differentiable rendering, where strands are represented as \textit{chained hair Gaussians}.

The final output of this pipeline is a collection of approximately $150K$ hair strands, each explicitly represented as a polyline with $N_k=100$ points.
While appearance parameters are also estimated as side-outputs, they are not the focus of this work.
Being prior-free, our method is designed to capture subject-specific details such as flying strands beyond the coverage of existing datasets.
Meanwhile, our method also strives to maintain the physical correctness of the hair geometry, including smoothness and scalp-connectivity. 
The estimated geometry is ready to be used in downstream pipelines such as physically based rendering, animation, and editing tasks.

\subsection{Data Acquisition and Preparation}
\label{subsec:overview:data}
We collect input data using our multi-camera system with $64$ cameras at $4K$ resolution under uniform illumination.
All cameras are calibrated, synchronized, and arranged on a sphere centered around the subject.
Depending on specific hairstyles, there are typically around $50$ cameras that capture the hairs, where the diagonal size of hair bounding box ranges from $1.3K$ to $4.6K$ pixels.
For each view, we compute a semantic segmentation mask that categorizes each pixel as either background, hair, or body (\ie non-hair foreground).
To ensure robust segmentation, we employ multiple off-the-shelf models and derive the final pseudo ground-truth labels using a simple aggregation strategy, elaborated in \autoref{app:seg}.

Using all views, we apply the technique in \cite{guo2019relightables} to achieve a rough surface reconstruction of the subject.
We then dilate the mesh by $2$cm to ensure all hairs are encompassed.
This mesh, referred to as the \textit{outer mesh}, sets a hard outer boundary for the subject and hairs.
We also fit a parameteric head mesh model to the captured subject using dense facial landmarks.
The resulting mesh of the fitted head model, referred to as the \textit{inner mesh}, approximates the bald surface of the subject's head and serves as the basis for locating the hair scalp where all strands originate.
Finally, we derive a loose 3D bounding box of the hairs by projecting per-view hair segmentation onto the outer mesh,.
The outer mesh, inner mesh, and bounding box together define the hair volume on which the whole pipeline operates.
If the hairstyle involves a visible parting line, we optionally accept a 2D annotation of the line from a selected top-down view.
This straightforward step is the only manual one in the pipeline and takes less than a minute to complete.
\autoref{fig:inputs} illustrates the inputs required for our method.

After all inputs are prepared, our pipeline works fully automatic without any further human intervention.
We consistently apply the same pipeline with identical parameters for all results in this paper, from both our in-house captures and public datasets that cover diverse hairstyles.

\section{Neural Hair Volume}
\label{sec:volume}

\begin{figure}[t]
  \includegraphics[width=\linewidth]{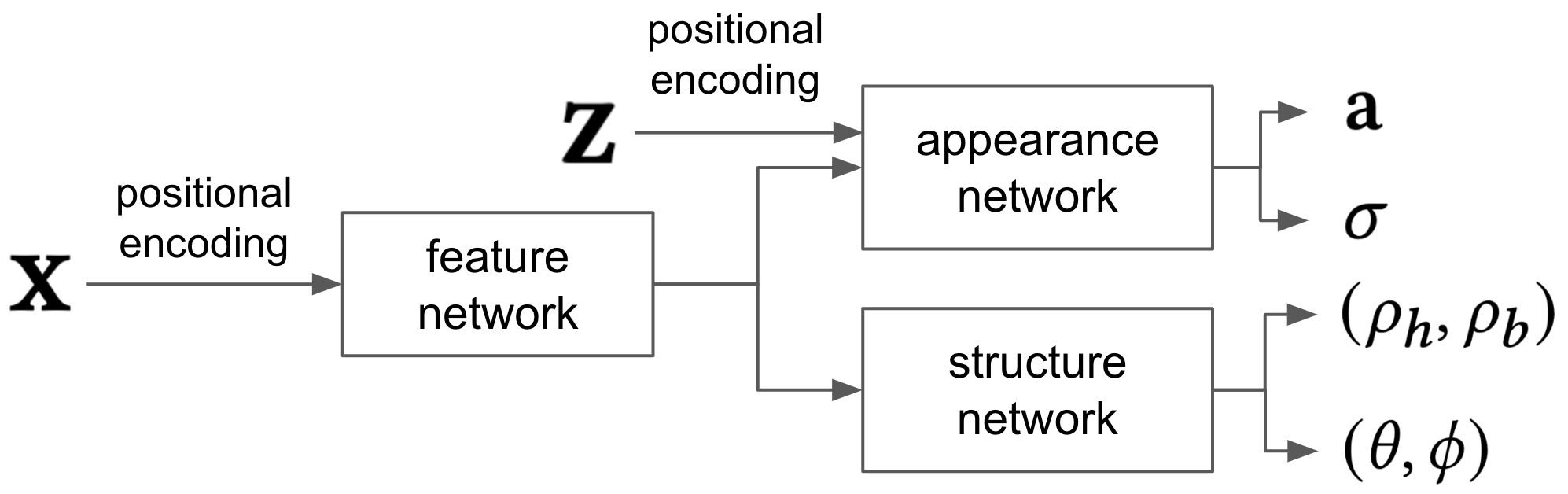}
  \caption{\textbf{The implicit hair volume network} comprises three sub-modules: the \textit{feature} network and \textit{appearance} network are used to estimate view-independent volume density $\sigma$ and view-dependent radiance $\rva$ from input position $\rvx$ and view direction $\rvz$, similar to NeRF; an additional \textit{structure} network is devised to estimate hair $\rho_h$ and body occupancy $\rho_b$ as well as 3D orientation $(\theta, \phi)$ in polar angles.}
  \label{fig:network}
\end{figure}

In the first stage, we build neural implicit fields to reconstruct the spatial orientation and occupancy of the subject's hairs, drawing parallels to neural radiance fields (NeRFs).
Our key contribution in this stage is to formulate a neural orientation field within the framework of volume rendering.

\subsection{Network Structure}

The implicit hair volume is formulated as an MLP network $\mathcal{V}$.
The input to $\mathcal{V}$ is a 3D position $\rvx \in \sR^3$, and the output includes volume density $\sigma \in [0, 1]$, hair occupancy $\rho_h \in [0, 1]$, body occupancy $\rho_b \in [0, 1]$ (refers collectively to the non-hair volume), and 3D hair orientation in polar angles $(\theta \in (0, \pi], \phi \in (0, \pi])$, all of which are view-independent.
Note that the polar angles are defined on a hemisphere because they are undirectional, \ie "lines" instead of "rays".
During training, we additionally feed $\mathcal{V}$ with the view direction vector $\rvz \in \sR^3$ and receive the view-dependent radiance color $\rva \in \sR^3$, similar to the vanilla NeRF.

As illustrated in \autoref{fig:network}, our model architecture comprises three sub-networks.
It begins with a shared \textit{feature network} that employs positional encoding to map an input 3D position $\rvx$ into a high-dimensional vector.
Subsequently, an \textit{appearance network} estimates view-independent volume density $\sigma$ and view-dependent color $\rva$ from the encoded position vector and additional view-direction input $\rvz$.
In parallel, the feature network branches into a \textit{structure network} that estimates 3D hair orientation $(\theta, \phi)$ and the occupancy values $(\rho_h, \rho_b)$ of hair and head.
In the following, we focus on the \textit{structure network} that is specifically devised for our task, while the feature- and appearance-networks are identical to the original NeRF.

\begin{figure}[t]
  \includegraphics[width=\linewidth]{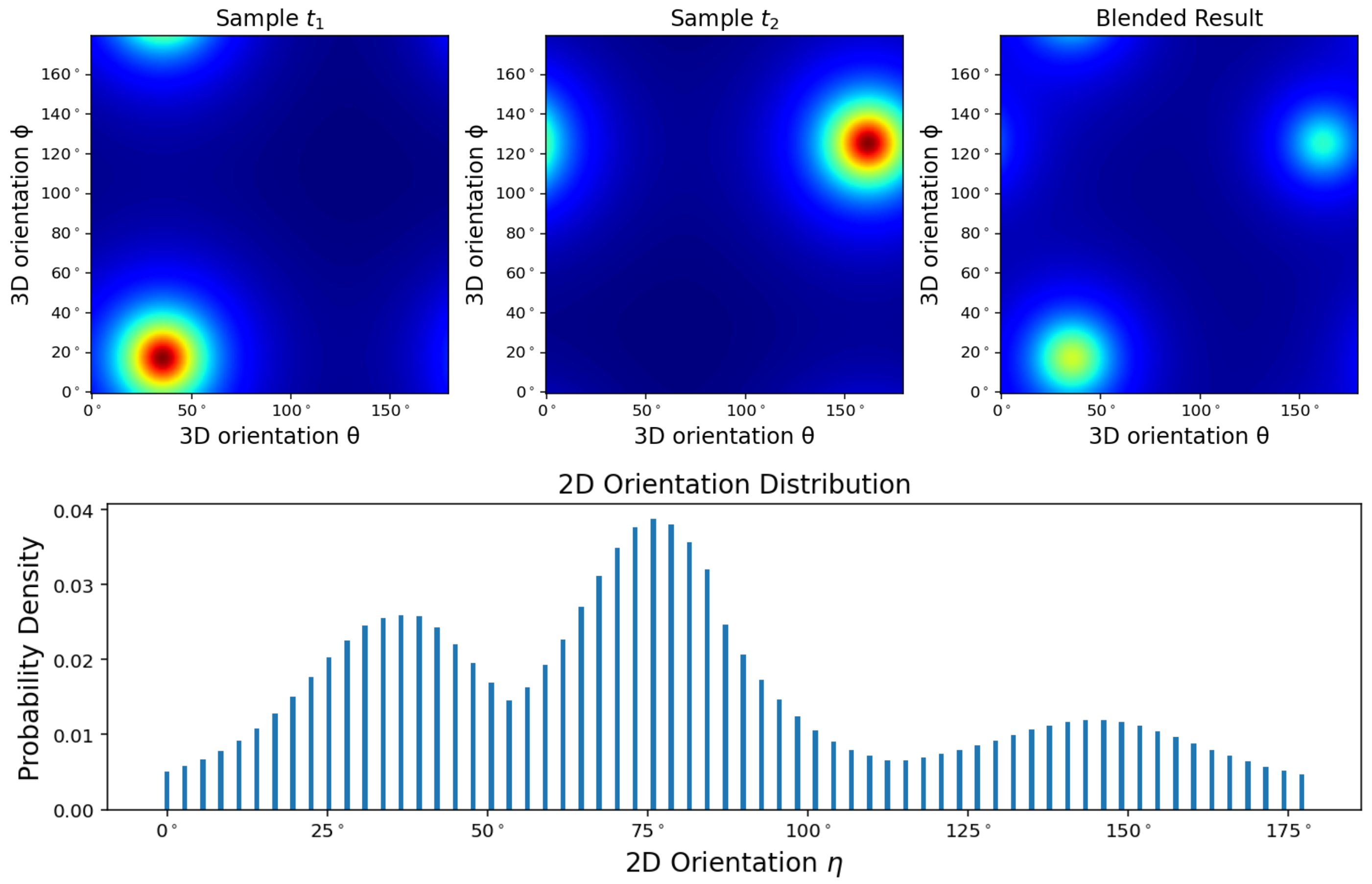}
  \caption{\textbf{Visualization of 3D orientation rendering and projection.} We take two exemplary samples $t_1$ and $t_2$, characterized by 3D orientations of $(0.1\pi, 0.2\pi)$ and $(0.7\pi, 0.9\pi)$ in polar angles. Both samples reside on the same ray with transmittances of $0.6$ and $1.0$, where $t_1$ is closer to the camera. In the top row, from left to right, we show the expanded 3D orientation distributions of $t_1$, $t_2$, and their blended integration. In the bottom row, we illustrate the 2D orientation distribution after projecting their integrated 3D orientations using an exemplary camera matrix, detailed in \autoref{eq:proj}.}
  \label{fig:alpha_blendering}
\end{figure}

\subsection{Neural Orientation Field}

\paragraph{Volume rendering of 3D Orientations}

The volumetric orientation plays a crucial role in defining the 3D hair structure.
Existing methods either optimize an explicit direction field with limited resolution \cite{wu24monohair}, or employ neural networks to predict the volumetric orientation in its entirety, which often leads to oversmoothed results \cite{saito183d}.
In contrast, our novel approach optimizes a neural orientation field that estimates 3D orientations without explicit limitation in resolution.
To construct this neural orientation field, we introduce a new formulation that ``renders'' 3D orientations within the volume rendering paradigm.

Volume rendering of 3D orientations is not as trivial as radiance.
Directly applying $\alpha$-blending to the polar angles is conceptually wrong.
For example, if a ray passes through two different hair strands whose orientations are $(\pi, 0)$ and $(0, \pi)$, assuming the transparency of the front hair is $0.5$, then the accumulated orientation becomes $(\pi/2, \pi/2)$, which is different from either hair and essentially smooths the distinct hairs into the same orientation.
The fundamental reason is that different orientations cannot be naively added together.
To accumulate the orientations of different hairs along a ray, we need to keep track of all angles.

To this end, we propose to expand a single 3D orientation, represented as polar angles, into a distribution, and perform $\alpha$-blending on the distributions.
Formally, for a 3D position $\rvx$, whose polar angles are $(\theta_\rvx, \phi_\rvx)$, we construct its distribution of 3D orientations $\mathcal{H}_\rvx$ by using a predefined kernel function as its probability density function (PDF) $h_{\rvx}$:
\begin{equation}
    h_\rvx(\theta, \phi) = \frac{1}{C_\rvx} h'_\rvx(\theta, \phi)
\end{equation}
\begin{equation}
    h'_\rvx(\theta, \phi) = \frac{1}{\beta(||\theta - \theta_\rvx||^2 + ||\phi - \phi_\rvx||^2) + \delta}
\end{equation}
\begin{equation}
    C_\rvx =  \iint_{0}^{\pi} h'_\rvx(\theta, \phi)\diff \theta \diff \phi \mathrm{.}
\end{equation}
Intuitively, $h(\cdot)$ is inversely proportional to the squared distance from an arbitrary angle $(\theta, \phi)$ to the ``center'' orientation $(\theta_\rvx, \phi_\rvx)$, with a scaling factor $\beta$, a damping factor $\delta$, and a divisor $C_\rvx$ that normalizes the integral to be 1.
We empirically found that this inverse-proportional function performs better than Gaussian kernels.
In practice, we furthermore consider the periodic and undirectional characteristics of orientations, replacing $h'$ with the more precise form $h''$:
\begin{equation}
    h''_\rvx(\theta, \phi) = \sum_{i=-1}^{1} \sum_{j=-1}^{1} \frac{1}{\beta(||\theta - \theta_0 + i\pi||^2 + ||\phi - \phi_0 + j\pi||^2) + \delta} \mathrm{.}
\end{equation}
The expanded distributions are illustrated in the top row of \autoref{fig:alpha_blendering}.

Based on the distribution formulation $\mathcal{H}_\rvx$, we can compute the accumulated 3D distribution $\mathcal{G}_r$ along an arbitrary ray $r$ with the following PDF:
\begin{equation}
    g_{r}(\theta, \phi) = \int_{t_n}^{t_f} T(t)\sigma\big(r(t)\big)h_{r(t)}(\theta, \phi)\diff t
\end{equation}
\begin{equation}
    T(t) = \mathrm{exp}\left(-\int_{t_n}^{t}\sigma(r(\alpha))\diff \alpha \right) \mathrm{.}
\end{equation}
These equations are derived from the classical volume rendering formulation, where $t$ is the depth value along the ray $r(t) =\rvo + t\rvq$ that originates from $\rvo$ with direction $\rvq$, $T(t)$ denotes the accumulated transparency along the ray, and $t_n, t_f$ are near and far planes.

In the actual implementation, we quantize the continuous integrals into discrete bins.
The range $(0, \pi]$ is divided into 64 bins such that the orientation distributions can be approximated by vectors of $64 \times 64$ dimensions, and the accumulation is performed for each dimension individually.
Accordingly, we set the scaling factor $\beta = (64/\pi)^2$ and $\delta = 0.01$.

\begin{figure}[t]
  \includegraphics[width=\linewidth]{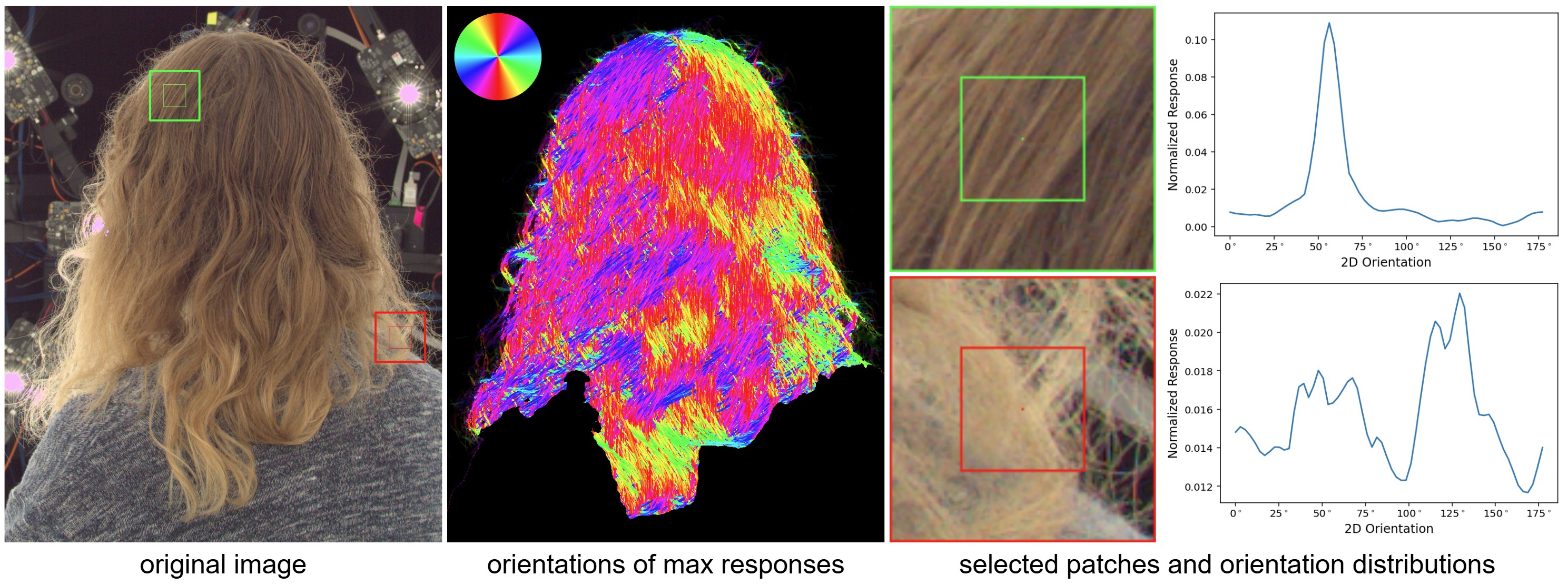}
  \caption{\textbf{Visualization of orientation distributions.} We start with an input image on the left, where we apply orientation filters and visualize the maximum responses in the middle. On the right, we select two exemplary patches (outlined by green and red rectangles on the left-hand side of the respective row), where the inner rectangles' sizes equal to the kernel radius. On the right-hand side, we illustrate the orientational distribution for each patch. For the green patch in the top row, there is a single sharp peak, indicating that most strands share the same direction. In contrast, the red patch in the bottom row shows a strong peak near $135^{\circ}$ and a secondary, broader peak from $30^{\circ}$ to $70^{\circ}$. The higher peak represents the thicker hairs that fill the lower-left half of the patch, while the other peak corresponds to the strands entering from the top-right corner. Merely keeping the maximum responses (middle) will omit these critical structural details.}
  \label{fig:distributions}
\end{figure}

\paragraph{Supervision with 2D Orientations}

Now we explain how to supervise accumulated 3D orientation distributions $\mathcal{G}$ using multi-view images.
In previous works, convolution with a bank of oriented filters was extensively used to estimate 2D orientation fields on hair images \cite{wu24monohair,paris04capture}, where the 2D orientation of each pixel is simply assigned as the angle of the filter that activates the maximum response.
However, we argue that representing a pixel's 2D orientation with a single value is insufficient.
First, a filter's receptive field may cover multiple hair strands with different orientations.
Second, several distinct hair strands may overlay and pass through the same pixel.
In both cases, only retaining a single angle will discard the valuable structural information of all other hairs, especially for challenging areas with high local inconsistency.
For example, \autoref{fig:distributions} inspects the responses of all filters at two patches.
When strand orientations within a patch are locally varied (as highlighted by the red rectangle in the second row), using the maximum response smooths distinct strands into the same orientation and leads to flat reconstructions.

To better preserve the structural information, we propose to maintain the responses of all filters, which naturally form a distribution of 2D orientations.
We use this 2D orientation distribution to supervise the neural orientation field, as it effectively aligns with our formulation that renders 3D orientations as a distribution.
Specifically, for each view with a known camera pose, we project the 3D orientation distribution $\mathcal{G}$ of each ray into a distribution of 2D orientations $\mathcal{F}$ with the following PDF:
\begin{equation}
\label{eq:proj}
    f(\eta) = \frac{1}{C_{\eta}} \max_{(\theta, \phi) \in u}h(\theta, \phi)
\end{equation}
where $\eta \in (0, \pi]$ is the 2D orientation, $u$ is a plane spanned by all $(\theta, \phi)$ pairs whose projection on the image plane is angle $\eta$, and $C_{\eta}$ normalizes the integeral to be 1.
Then, we define the loss function for the neural orientation field as
\begin{equation}
    \mathcal{L}_{\mathrm{ori}} = \int_{0}^{\pi} ||f(\eta) - \bar{f}(\eta)||^2 \diff \eta
\end{equation}
where $\bar{f}(\eta)$ is the normalized response of the orientation filter at angle $\eta$.

In our quantized implementation, $\eta$ is discretized into $64$ values.
We use $64$ Gabor filters to convolve over the grayscale image and store the responses of all filters.
The correspondence between 2D orientation $\eta$, plane $u$, and 3D orientation $(\theta, \phi)$ can be easily enumerated.
An illustration of a projected 2D orientation distribution can be found in the bottom row of \autoref{fig:alpha_blendering}.

\subsection{Neural Occupancy Field}

Our implicit hair volume further establishes neural occupancy fields by predicting hair occupancy value $\rho_h$ and body occupancy value $\rho_b$ at any given position.
The continuous values of hair occupancy naturally align with the fact that hairs are semi-transparent in images.
The occupancy values $\rho_*$ are accumulated using the standard volume rendering formula to give per-pixel labels $\psi_{*}$ and supervised by pseudo ground truth (GT) segmentation labels $\bar{\psi}_{*}$:
\begin{equation}
    \label{eq:l_occ}
    \mathcal{L}_\mathrm{occ} = ||\psi_h - \bar{\psi}_h||^2 +||\psi_b - \bar{\psi}_b||^2 .
\end{equation}
Notably, the GT masks do not need to be perfect.
We find the segmentations estimated by our method finally outperform GT, since they implicitly integrate multi-view information.

\subsection{Training Strategy}

The model undergoes a two-phase training process.
Initially, only the feature and appearance networks are trained with the conventional L2 photometric loss.
In the subsequent phase, the structure network is trained alone with loss $100 \mathcal{L}_\mathrm{ori} + 0.02 \mathcal{L}_\mathrm{occ}$, and the other two modules are frozen.
This two-phase training strategy enhances stability and convergence.
In our experiments, we find that only supervising radiance in the first phase mitigates the risk of $\sigma$ predictions being contaminated by the relatively noisy 2D semantic and orientation labels.
Following \cite{sarkar2023litnerf}, we utilize the reconstructed outer mesh to decide the depth sampling range of the rays.
This ensures that the model focuses exclusively on the hair volume.

\section{Volumetric Hair Tracing}
\label{sec:tracing}

Once the hair volume is established, we extract hair strands by tracing within the volume using the inferred volumetric orientation and occupancy with forward Newton method \cite{paris08hair,chai13dynamic,kuang22deepmvshair}.
Specifically, at timestep $k$, each strand is extended by a fixed length $l = 3\mathrm{mm}$ to a new point $\rvv_k=\rvv_{k-1}+l \cdot \mathrm{norm}(\rvm_k)$, where the growing direction $\rvm_k$ before normalization is calculated as:
\begin{equation}
\begin{split}
    \rvm_k=&\gamma\cdot\rvm_{k-1}+\\&(1-\gamma)\cdot\big(\sign(\rvg\cdot\rvm_{k-1})\cdot\rvg-\lambda\min(\rvn\cdot\rvm_{k-1},0)\cdot\rvn\big).
\end{split}
\end{equation}

In this formula, $\rvg$ represents the predicted 3D orientation derived from polar angles.
We disambiguate it to the direction most closely aligned with the previous strand direction $\rvm_{k-1}$, determined by the sign function $\sign(\cdot)$ applied to the dot product.
Additionally, an inertia term controlled by the factor $\gamma$ encourages smooth growth transitions to avoid abrupt changes caused by potential outliers.
A surface repulsion term controlled by factor $\lambda$ is used to push strands away from the head according to the surface normal $\rvn$ of the inner mesh.
In our implementation, we set $\gamma=0.6$ and $\lambda=f/f_d$, where $f$ is the current penetration distance between $\rvv_{k-1}$ and the inner mesh ($0$ if not penetrating), and $f_d = 5$mm is a constant penetration threshold.

We initialize tracing from seed points uniformly sampled within the bounding box volume between the inner and the outer mesh.
These seeds are organized into a priority queue, weighted by the product of volume density and hair occupancy, $\sigma \cdot \rho_h$. We also dynamically deprioritize seeds in close proximity to newly traced strands to promote volumetric uniformity.

During tracing, we monitor a health value for each strand and cease tracing when this value drops to $0$.
At each step, the health is reduced if:
1) the current vertex has low volume density $\sigma$ or hair occupancy $\rho_h$;
2) the vertex goes outside the outer mesh or the bounding box.
Strands traced from seeds in this step are referred to as \textit{volume hairs}, which are not guaranteed to connect to the scalp.

The primary challenge with volume hairs not being scalp-rooted is unreliable structural information near the scalp due to severe occlusion. To address this, we trace an additional set of \textit{scalp hairs} to serve as a bridge between the floating volume hairs and the scalp surface. Scalp hairs are initiated by sampling seeds on the scalp region of the inner mesh, with their initial growth directions set to the normals of the scalp. These hairs are then grown similarly to volume hairs.

Once all scalp hairs are obtained, for each volume hair, we randomly select a nearby scalp hair and grow the volume hair backward along this scalp hair to the scalp.
This process is able to connect most strands (typically more than $99\%$) to the scalp.
Any volume hairs that fail to connect to the scalp are eventually discarded.
Additionally, if a parting line is annotated for the hairstyle, we remove all hairs crossing it as a refinement step.

The final output of this stage is a collection of $N_s$ strands $\mathcal{S} = \{\rvs_1, \rvs_2, ..., \rvs_{N_s}\}$.
We resample each strand to $N_k = 100$ vertices, \ie $\rvs_i \in \sR^{N_k \times 3}$.
Although our tracing algorithm can theoretically generate an arbitrary number of hairs, we target $25K$ scalp hairs and $125K$ volume hairs in all cases.

\section{Gaussian-Based Strand Optimization}
\label{sec:3dgs}

After navigating through the aforementioned pipeline from 2D orientation estimation to implicit hair volume prediction and ultimately to strand tracing, we observe a gradual loss of structural information.
In this final stage, we seek direct supervision from the original images to recuperate the lost fine details, ensuring a proper match to the captured imagery.

To achieve this, we adopt the image-based differentiable rendering framework of 3D Gaussian spatting (3DGS) \cite{kerbl20233d} to optimize the reconstructed hairs using photometric losses.
Our method introduces a novel chained hair Gaussian formulation that constrains the relationships among Gaussians along each strand, aligning with the inherent geometric nature of hair.
While the concurrent work of GaussianHair \cite{luo2024gaussianhair} also proposes to use cylinderal Gaussians as hair proxies, our formulation avoids hallucination effects with rigorously designed constraints.

\subsection{Formulation of Chained Hair Gaussians}

In contrast to the vanilla 3DGS framework, our optimization targets are the parameters of hair geometry, rather than the shape and appearance parameters of individual Gaussians.
We now describe the conversion from strand geometry to chained Gaussians in our representation, which correlates with strand parameters while remaining compatible with the Gaussian splatting framework.

We define the elementary unit of strands as line segments.
For a strand of $N_k$ vertices, we denote the segment between vertex $\rvv_i$ and $\rvv_{i + 1}$ by $\rvp_i$, characterized by the following parameters: head vertex $\rvv_i$, tail vertex $\rvv_{i+1}$, diameter $d$, opacity $o$, and spherical harmonics (SH) coefficients $\rvr$.

In our chained Gaussian representation, each segment $\rvp_i$ is approximated by a Gaussian centered at the midpoint $(\rvv_i + \rvv_{i + 1}) / 2$. The covariance matrix $C$ of this Gaussian is expressed as:
\begin{equation}
    \label{eq:cov}
    C = E D D^T E^T.
\end{equation}
Here, $E = [\rve_i, \rve'_i, \rve''_i]^T$ represents the principle axes of the Gaussian, where $\rve_i$ is the unit direction vector of the segment $\rvv_{i+1} - \rvv_i$, and $\rve'_i$ and $\rve''_i$ are two orthogonal unit vectors to $\rve_i$.
The matrix $D = \mathrm{diag}[\tau_l, \tau_d, \tau_d]$ contains scales of the axes, with $\tau_l = ||v_{i+1} - v_{i}|| / 2$ and $\tau_d = d / 2$ being the axial and radial scales, respectively.

Following this conversion, each strand is transformed into a chain of thin Gaussians, suitable for rendering via Gaussian splatting.
At the conclusion of optimization, we typically manage $50K$ strands, equivalent to $5M$ Gaussians, and render only one-third of them picked randomly due to limited memory capacity.

In addition to hair Gaussians, we also incorporate auxiliary Gaussians to model the non-hair foreground, serving as proxies for occlusion.
These Gaussians, referred to as \textit{body Gaussians}, are anchored at the vertices of the inner mesh and modeled as discs with optimizable radii $w$.
The orientation of each body Gaussian disc is fixed, aligned with the normal of the corresponding vertex. The thickness is set to $0.001$mm.
The covariance matrix of each disc is calculated as described in \autoref{eq:cov}, treating each Gaussian as a short, thick disc covering the surface.
Given our focus on hair, we differentiate non-hair pixels in each image by painting them a distinct color from hair, such as green.
Accordingly, the body Gaussians are initialized to the same green color.

\subsection{Geometry Parameters}

Instead of directly optimizing the positions of strand vertices, we optimize a low-dimensional latent vector for each strand. This improves training efficiency while also serving as an effective regularizer, preventing exaggerated strand geometry such as unnatural sharp turns, which occurs with per-vertex optimizations.

To build such a strand latent space, previous efforts typically leverage generic prior models based on synthetic curves.
However, we argue that building such a comprehensive space is impractical due to the high variability of real-world hairs and the significant domain gap.
Instead, we construct this strand latent space in a self-supervised manner, relying solely on the initial strands of the specific subject.
This approach aligns with the intuition that the hairs of the same individual should share statistical similarities.

Specifically, for each subject, we train a strand variational autoencoder (\textit{strand-VAE}) that encodes a latent code $\rvl \in \sR^{128}$ from root-relative vertex positions $\rvs' \in \sR^{(N_k - 1) \times 3}$.
While the strand-VAE is a vanilla MLP network, it works better than more complex generic models, such as \cite{zhou2023groomgen, rosu2022neural}, since training a subject-specific latent prior is a much easier task.
A less complex model structure also simplfies optimization.
The strand-VAE is initialized from scratch and trained only with the traced hair strands.
In our experimental set-up we show that the quality and diversity of the data is sufficiently good to serve as a dataset.

\subsection{Appearance Parameters}

While the high degree of freedom (DoF) associated with Gaussian parameters enables effective reproduction of appearance, it also introduces severe hallucination effects when precise geometry is desired, particularly when it comes to the intricate structure of hairs.
As long as the volume is reasonably filled, even if the strand geometry is inaccurate, the high DoF of color parameters can fabricate the appearance to simulate shading that minimizes photometric errors, without the support from proper geometry.
This necessitates additional constraints in hair parameterization to ensure that improvements in appearance actually result from enhancements in geometry.

To limit the per-strand appearance DoF, we propose the following simplifications to the hair appearance parameters:
\begin{itemize}
    \item \textit{View variations of color}: We eliminate view-dependent components of color by reducing the SH degree to zero, which is non-essential for our method that prioritizes geometry optimization. This adjustment prevents the potential misuse of view-dependent effects.
    \item \textit{Spatial variations of color}: Instead of maintaining a color parameter for each strand segment, we optimize the color for only 8 segments (referred to as \textit{anchors}) uniformly distributed along the strand. The color for other segments is derived via piecewise linear interpolation.
    \item \textit{Segment diameters}: Similar to color, we also parameterize segment diameters using these 8 anchors.
    \item \textit{Opacity}: We restrict each strand to $2$ opacity values: $o_1$ for the first $N_k - N_t - 1$ segments starting from the root, and $o_2$ for the final $N_t = 8$ segments, recognizing that the tails of strands tend to be more transparent.
\end{itemize}

In conclusion, for each strand, we optimize the following parameters: strand latent vector $\rvl \in \sR^{128}$, anchor diameters $\rvd \in \sR^{8}$, anchor colors $\rvr \in \sR^{8\times3}$, and opacity $(o_1, o_2) \in \sR^2$.
By comparison, the vanilla 3DGS setting would involve nearly $1400$ optimizable parameters per strand, which is approximately $8$ times more than the $162$ parameters of our streamlined formulation.

\subsection{Adaptive Control of Hair Gaussians}

During optimization, we adaptively control the strand distribution by periodically employing heuristic-based actions including splitting and pruning.
The Gaussian optimization starts with $30K$ initial strands uniformly sampled from the traced hairs, and gradually propogates them to $50K$ in the end. This adaptive control leads to a more natual hair arrangement.

\paragraph{Splitting}
With the introduction of diameter parameters, individual strands are allowed to grow thicker and split into multiple new strands where necessary.
For each strand $s_i$ with $N_k - 1$ segments, given its per-segment diameters $d_{i, j}$ and opacities $o_{i, j}$ for the $j$-th vertex, we compute the per-strand split score $\omega_i$ the following way:
\begin{equation}
\label{eq:split}
    \omega_i = \frac{\hat{\omega_i}}{\frac{1}{N_s} \sum_{i = 1}^{N_s}\hat{\omega}_i}, \quad \hat{\omega}_i = \sum_{j=1}^{N_k - 1} d_{i,j} \cdot o_{i,j} \mathrm{.}
\end{equation}
where $N_s$ is the number of strands at the time of splitting.
Each strand is then split into $\lceil \omega_i \rceil$ new strands, whose vertices are generated by randomly displacing the original positions within its diameter.

This splitting operation serves two purposes in improving optimization quality.
First, it dynamically adjusts hair density to achieve a more uniform distribution that aligns with the image data.
Second, it creates flyaway strands to capture details that are missing in the initial hair tracing.

\paragraph{Pruning}
Besides splitting, we regularly perform clean-up operations, including pruning and cutting, to preserve only meaningful strands and segments.
Throughout the optimization, certain strands may migrate outside the designated hair volume, thus becoming transparent or blending into the color of the background. These strands, practically invisible and uninteresting, should be removed completely.
Additionally, some strands can grow excessively long into void spaces, necessitating the cutting of their tails instead of the entire strands.

In our method, we identify and prune those invisible strands based on their opacity and color.
We periodically remove strands whose average vertex opacity falls below a threshold of $0.1$.
Furthermore, we calculate the average hair color $\bar{\rvr}^\mathrm{h}$ of all strands in CIELAB color space, pruning each strand $\rvs_i$ if its average color $\rvr_i$ is closer to the background color $\rvr^\mathrm{b}$ than to the average hair color $\bar{\rvr}^\mathrm{h}$.
Similar checks are conducted for strand vertices individually.
We remove consecutive invisible vertices from strand tails until the first visible vertex, according to the aforementioned criteria.
While the invisible vertices in the middle of strands are not affected, in practice they are very few and do not harm the reconstruction quality.

\subsection{Training Objectives}

The primary loss during optimization is the L2 photometric distance between rendered images and reference images, denoted as $\mathcal{L}_\mathrm{i}$.
In addition to this, we introduce the following regularization terms.
For the sake of simplicity, we omit per-strand subscripts.
The final loss terms are computed as the mean over all strands.

\paragraph{Volume Guidance Term}
We reuse the implicit hair volume model from the first stage to provide additional 3D guidance:
\begin{equation}
    \mathcal{L}_\mathrm{n} = \frac{1}{N_k - 1} \sum_{i=1}^{N_k-1} \min(||\rve_i - \rvg_i||, ||\rve_i + \rvg_i||),
\end{equation}
where $\rve_i$ is the direction of the hair segment, and $\rvg_i$ is the undirectional 3D orientation prediction at $(\rvv_{i + 1} + \rvv_{i}) / 2$.
This term helps to regularize strands that do not receive adequate gradient information from image pixels.

\paragraph{Penetration Prevention Term}
We introduce a penetration loss to prevent hairs from growing inside the inner mesh:
\begin{equation}
    \mathcal{L}_\mathrm{p} = \frac{1}{N_k} \sum_{i=1}^{N_k} ||\rvv_i - \tilde{\rvv}_i||^2,
\end{equation}
where $\tilde{\rvv}_i$ is the nearest point on the inner mesh surface to $\rvv_i$.
This term is applied only if $\rvv_i$ is already located inside the mesh.

\paragraph{Heursitic Terms}
Finally, we incorporate the following heuristic-based terms:
\begin{itemize}
    \item \textit{Diameter term}: $\mathcal{L}_d = \sum_{i=1}^{N_k - 1} |d_i| / (N_k - 1)$ to encourage strands to be thin and sharp;
    \item \textit{Latent regularization term}: $\mathcal{L}_\rvl = |\rvl - \hat{\rvl}|$ to regularize the strand's latent vector $\rvl$ towards its initial value $\hat{\rvl}$, obtained from the hair tracing stage;
    \item \textit{Body radius term}: $\mathcal{L}_b = \sum_{i=1}^{N_b} ||w_i - \hat{w}_i ||^2/N_b$, a regularization on the radii $w$ of body Gaussians, where $\hat{w}_i$ is the initial radius.
\end{itemize}

The overall training objective is thus:
\begin{equation}
    \mathcal{L} = \lambda_\mathrm{i}\mathcal{L}_\mathrm{i} + \lambda_\mathrm{n}\mathcal{L}_\mathrm{n} + \lambda_\mathrm{p}\mathcal{L}_\mathrm{p} + 
    \lambda_\mathrm{d}\mathcal{L}_\mathrm{d} +
    \lambda_\mathrm{\rvl}\mathcal{L}_\mathrm{\rvl} +
    \lambda_\mathrm{b}\mathcal{L}_\mathrm{b},
\end{equation}
where we set the weights as $\lambda_\mathrm{i} = 1$, $\lambda_\mathrm{n} = 1.0$, $\lambda_\mathrm{p} = 0.05$,  $\lambda_\mathrm{\rvl} = 1.0$, and $\lambda_\mathrm{b} = 1000$.
The weight for diameter regularization, $\lambda_\mathrm{d}$, starts at $1$ and is doubled after each strand splitting step.

\begin{figure*}[t]
  \includegraphics[width=\linewidth]{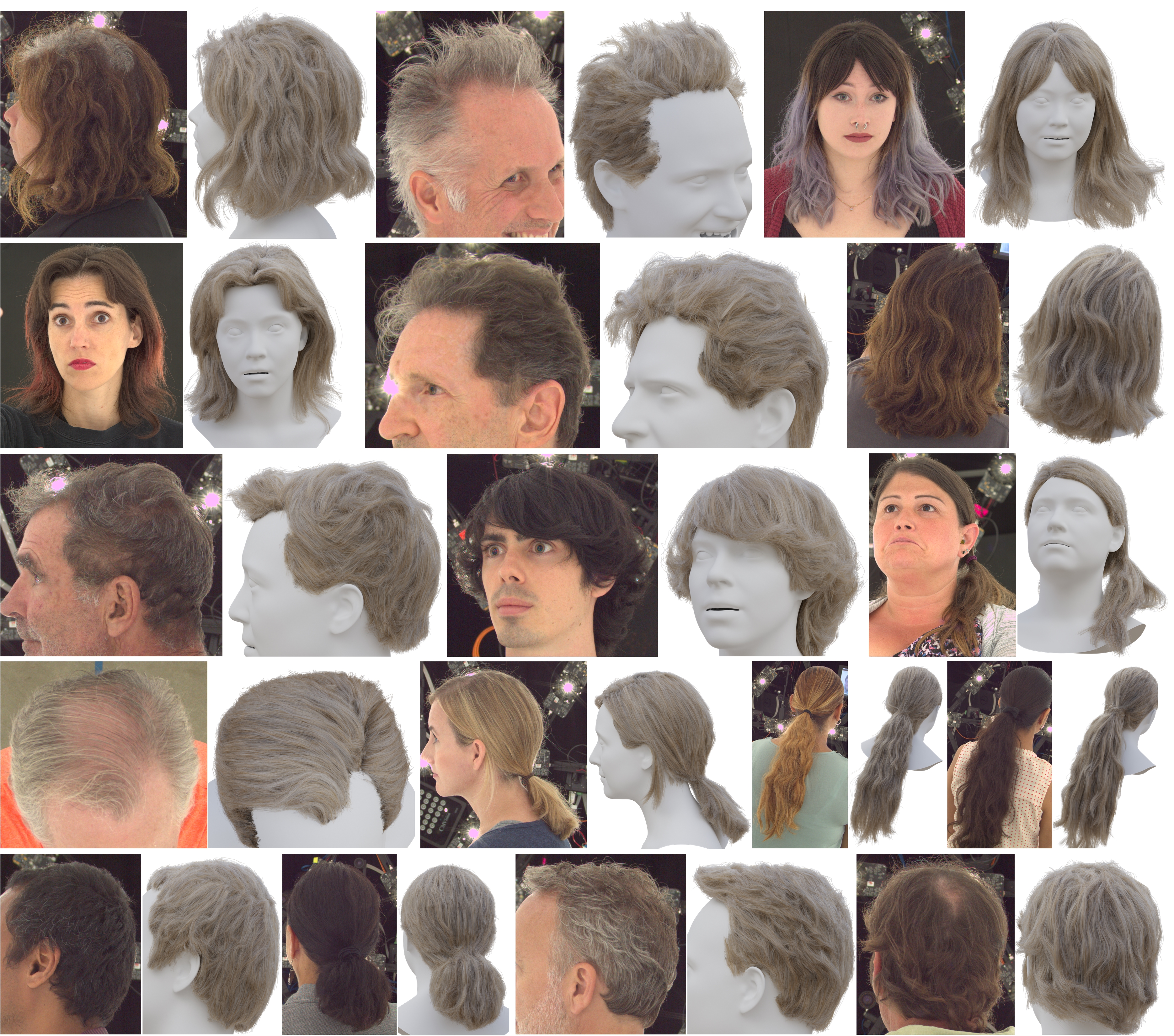}
  \caption{\textbf{Reconstruction results} on diverse hairstyles from short hairs to long ponytails, where personal features such as fringe, hairline, and clusters are faithfully captured. We use the same predefined material to better show geometric details.}
  \label{fig:major_results}
\end{figure*}

\begin{figure}[h]
  \includegraphics[width=\linewidth]{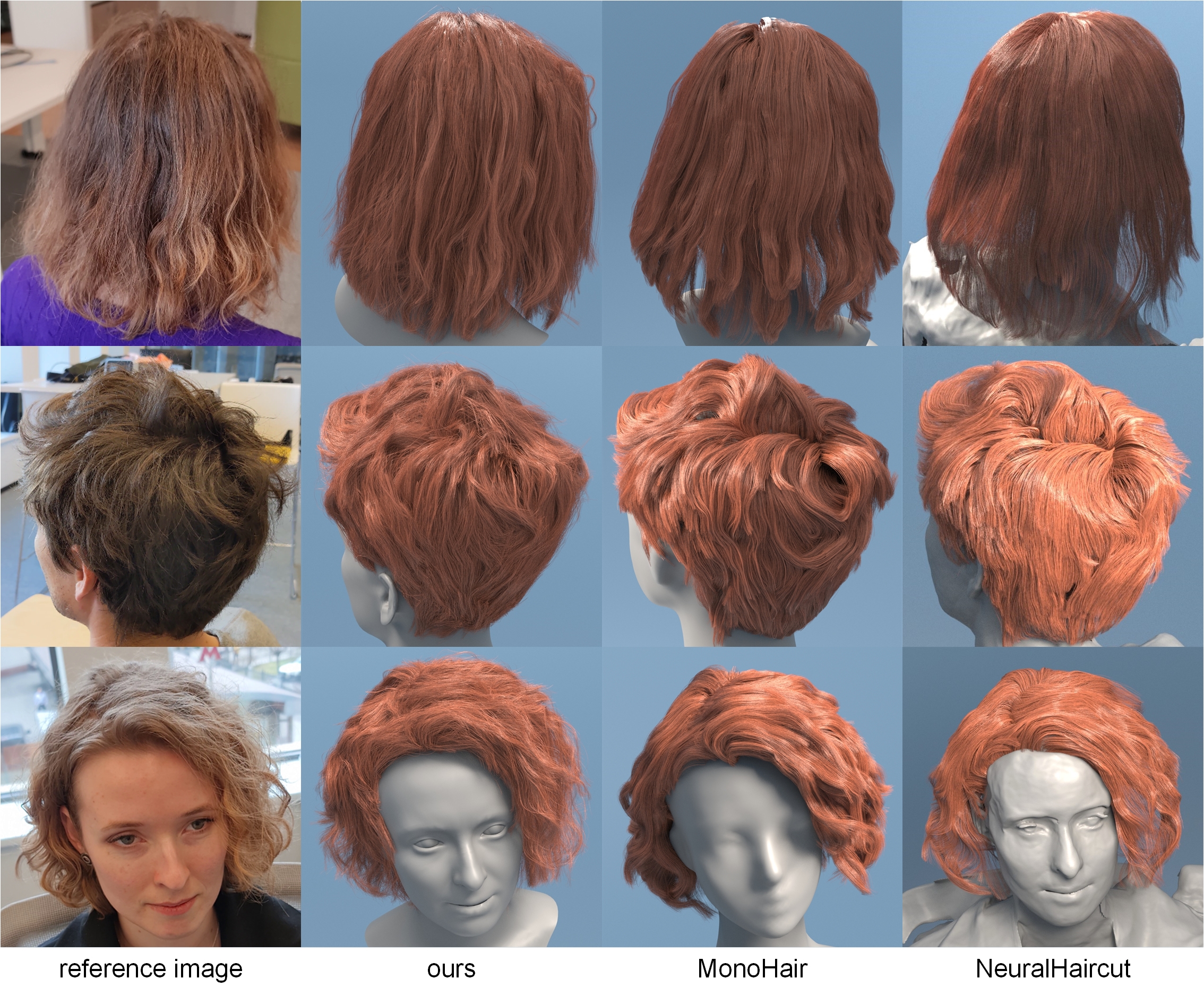}
  \caption{\textbf{Comparions with existing hair reconstruction methods.} We compare GroomCap with MonoHair \cite{wu24monohair} and Neural Haircut \cite{sklyarova23neural} on the in-the-wild NHC dataset, rendered with the same renderer. The rendering camera of NeuralHaircut results are manually adjusted to match the input.}
  \label{fig:comparison}
\end{figure}

\section{Experiments}

In this section, we first describe implementation details and external datasets used for experiments.
Then, we compare with previous works and validate the key design choices with in-depth ablation studies.
Finally, we discuss limitations and illustrate failure cases.

To thoroughly test the versatility and robustness of our method, we take $27$ subjects from our in-house capture dataset, covering a broad spectrum of different hairstyles.
All subjects are processed automatically using the same pipeline.
For all results presented in this paper, unless specified otherwise, we eliminate all hallucination effects by rendering the reconstructed geometry with predefined hair materials.
This ensures that all shading effects, such as highlights, shadows, and transparency, faithfully reflect the quality of the geometry.
We present all visualizations in high-resolution figures.
Readers are encouraged to zoom in on these figures for better details.
More illustrations and applications are available in the Appendix and supplementary video.

\subsection{Implementation}

For our implicit hair volume model, the feature network comprises $6$ fully-connected (FC) layers, the appearance network consists of $2$ FC layers, and the hair structure network includes $8$ FC layers, each having $1024$ hidden units.
We train both the feature and appearance networks for $1M$ steps, followed by an additional $1M$ steps for the hair structure network alone.
On average, the training process takes $28$ hours in total, using $16$ Google TPU v5.

For example, we currently render a full orientation distribution in the neural hair volume, leading to significantly increased computational overhead due to the large number of bins in the histogram. 

The volumetric hair tracing is completed in $1.5$ hours using a single NVIDIA A100 GPU.
The bottleneck lies in repeated queries to the implicit hair volume, which can be potentially boosted by precomputing an explicit volume at high resolution.

Our Gaussian-based optimization encompasses $15K$ steps, taking $1.5$ hours on $8$ NVIDIA H100 GPUs.
During the optimization, we execute hair splitting and pruning every $5K$ steps, increasing the number of strands from $30K$ to $50K$.
We further apply a final splitting step that increase the hairs to $150K$ by scaling the split score $\omega$ in \autoref{eq:split} by $3$.
The strand-VAE network comprises $6$ encoder and $4$ decoder layers, taking $2.5$ hours to train over $1M$ steps on a single NVIDIA A100 GPU.

Prioritizing reconstruction fidelity for production use cases, the entire pipeline of our method requires more computation than NeuralHaircut \cite{sklyarova23neural} (around $3$ days on a NVIDIA RTX 4090 GPU) and MonoHair \cite{wu24monohair} ($4-6$ hours on a NVIDIA RTX 3090 GPU). This is partially because that we consider a full orientation distribution represented by a large number of bins in the histogram. Therefore, our reconstruction preserve more details at the cost of longer processing time.

\subsection{Results}

In \autoref{fig:major_results}, we demonstrate our results on various hairstyles captured in our studio.
Being prior-free, our method can reconstruct diverse hairstyles that surpass the coverage of any existing dataset, capturing personal details such as hairlines, fringes, and clusters.
Our approach not only handles common medium-length hair, but also deals with short hairs and long ponytails using the same pipeline, which are rarely addressed in previous works -- short hairs pose challenges due to their messiness and inconsistent patterns, while long ponytails often exceed the capacity of explicit volumes utilized in prior methods.
Our method also performs well on dark and dyed hairs, a notable improvement over previous works that typically focus on brighter hair colors.

\subsection{Comparisons}

We extend the evaluation of our method to the public NeuralHaircut (NHC) dataset \cite{sklyarova23neural} and compare with previous works.
Unlike our main dataset that is captured in a more controlled setup, the NHC dataset comprises only videos captured in-the-wild using a hand-held smartphone. This setup introduces additional challenges, such as unknown camera poses, non-uniform lighting, and slight subject movements during capture.
To preprocess the NHC dataset automatically, we employ colmap \cite{schoenberger2016sfm} for each video to estimate camera poses. After this, we fit the inner meshes using the same parametric head model.
The outer mesh is approximated simply by a sphere, roughly enclosing the head and hair.
Furthermore, we do not utilize any parting-line annotations for this dataset.

In \autoref{fig:comparison} we compare our results with state-of-the-art multi-view hair reconstruction works MonoHair~\cite{wu24monohair} and Neural Haircut~\cite{sklyarova23neural} on the in-the-wild NHC dataset.
While our reconstructions on this dataset are inferior than our primary setting due to the imperfect inputs, they remain comparable with the concurrent work of MonoHair and outperforms the earlier work of NeuralHaircut.
Unlike these works, which are strongly regularized by prior models, our prior-free reconstructions offer better flexibility and yield more visually realistic outcomes.

\begin{figure}[h]
  \includegraphics[width=\linewidth]{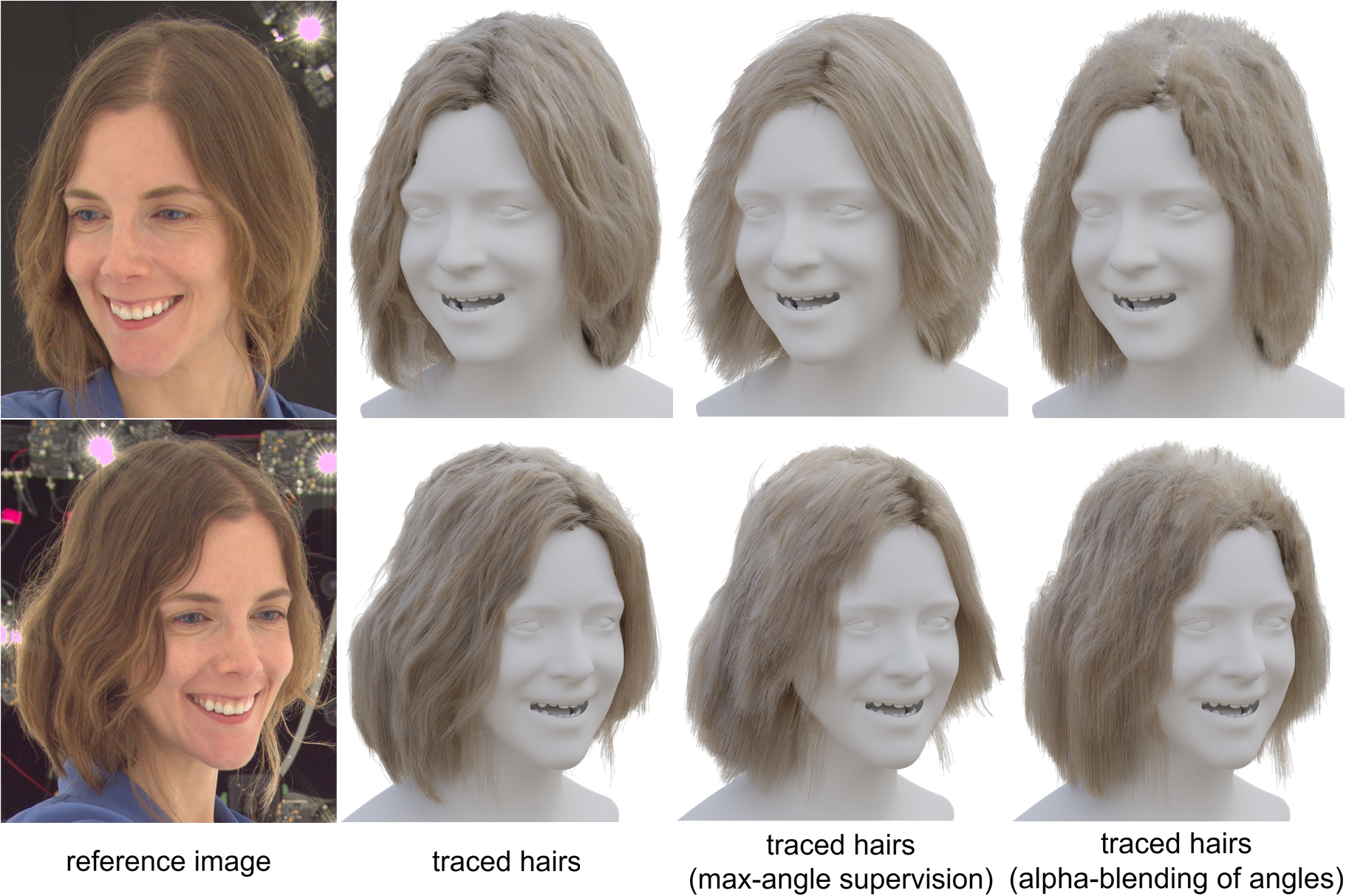}
  \caption{\textbf{Ablation studies for implicit hair volume.} We show strands traced from different hair volumes, including full method (second column), 2D supervision of maximum orientations without keeping the distribution (third column), and directly $\alpha$-blending 3D orientation angles without our rendering algorithm (fourth column). The results are either overly smoothed (third column) or contain incomplete and sparser strands (fourth column).}
  \label{fig:nerf_ablation}
\end{figure}

\begin{figure}[t]
  \includegraphics[width=\linewidth]{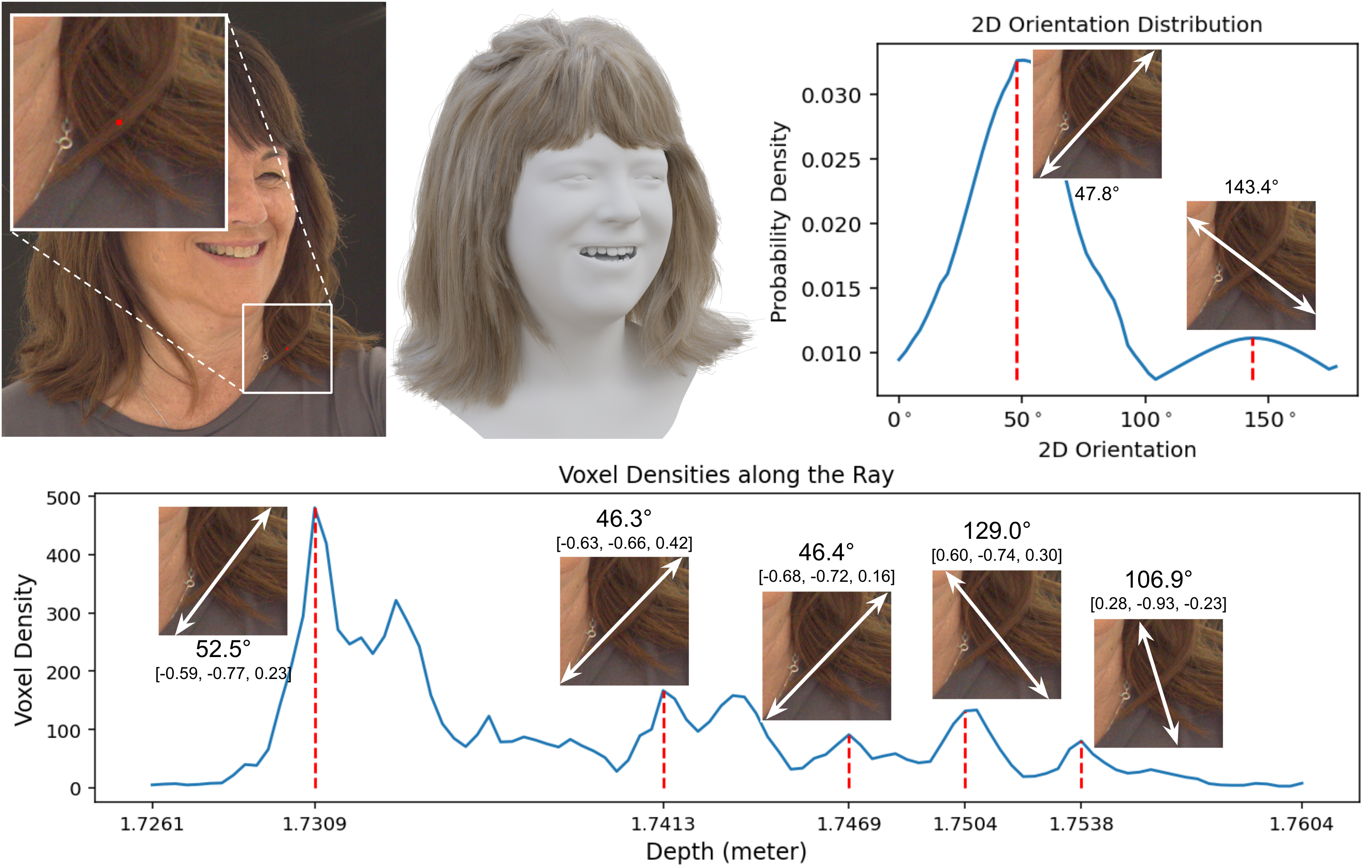}
  \caption{\textbf{Visualization of 3D orientation predections.} On an example subject, we show a reference view (top-left) and the corresponding hair reconstruction (top-middle). In the reference view, we highlight a sample patch (the white square) where two intersecting wisps are accruately captured in the output.
  In the lower part of this figure, we plot voxel densities along the ray path at the center of the patch using a line chart. For each density peak, we visualize the corresponding predicted 3D orientation by drawing an arrow over the patch. The first peak represents the front hair wisp with a 3D orientation in camera space of $[-0.59, -0.77, 0.23]$ and a 2D projection of $52.5^\circ$. Beginning at depth $1.75$m (the fourth peak), the ray intersects the back layer of hair, with 2D projections ranging from $107^\circ$ to $130^\circ$.
  At the top-right, we visualize the accumulated 2D orientation distribution along the same ray at the patch center, identifying two peaks. The first peak at $48^\circ$ correlates to the front hairs, while the second peak at $143^\circ$ corresponds to the hair at the back.}
  \label{fig:dist_3d}
\end{figure}

\begin{figure*}[t]
  \includegraphics[width=\linewidth]{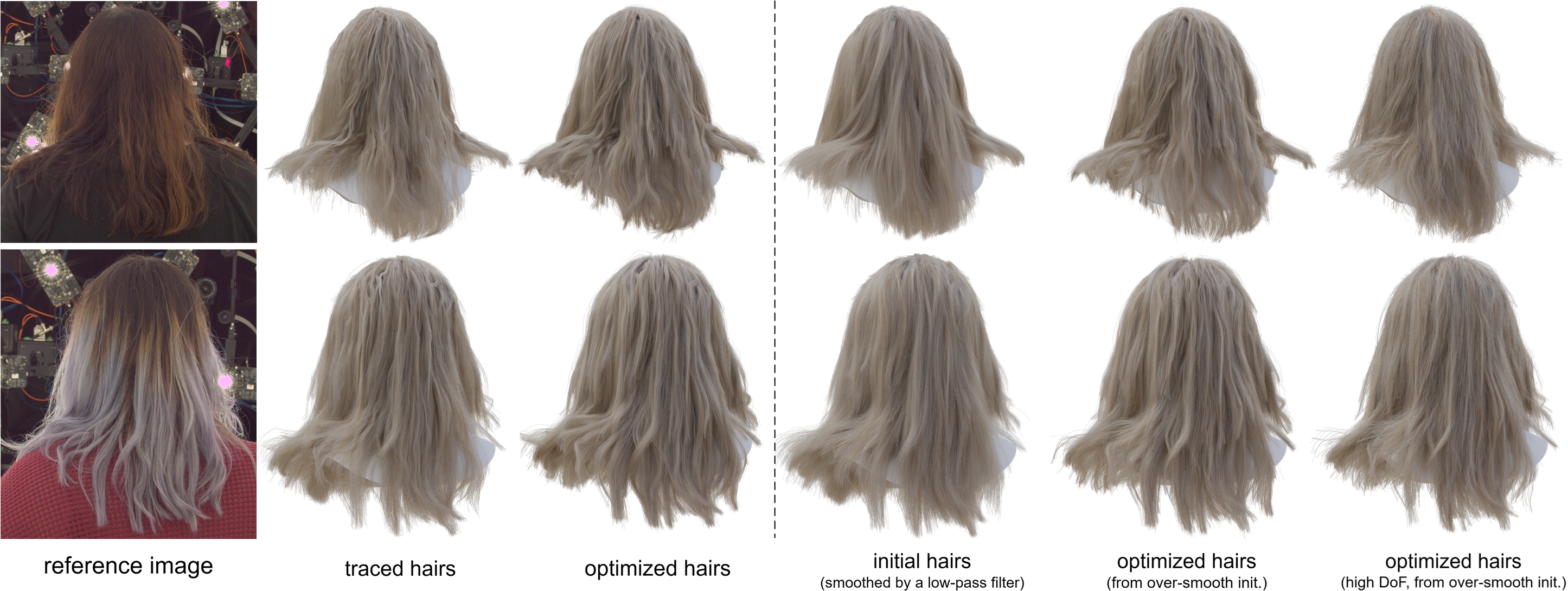}
  \caption{\textbf{Ablation studies for Gaussian-based hair optimization.} In the second and third columns, we show hair models before and after optimization, respectively. The optimization effectively consolidates the hair boundary and enhance overall smoothness.
  In the fourth column, we show an initial hair model that is intentionally smoothed from the traced hairs to better highlight the difference brought by optimization. The fifth column demonstrates that, even from this smoothed initial hair, the optimization is capable of faithfully recovering detailed features.
  However, as shown in the sixth column, keeping the high degree-of-freedom parameters of the vanilla 3DGS leads to flattened strands, which underscores the importance of our tailored Gaussian parameters.}
  \label{fig:gs3d_ablation_major}
\end{figure*}

\begin{figure}[h]
  \includegraphics[width=\linewidth]{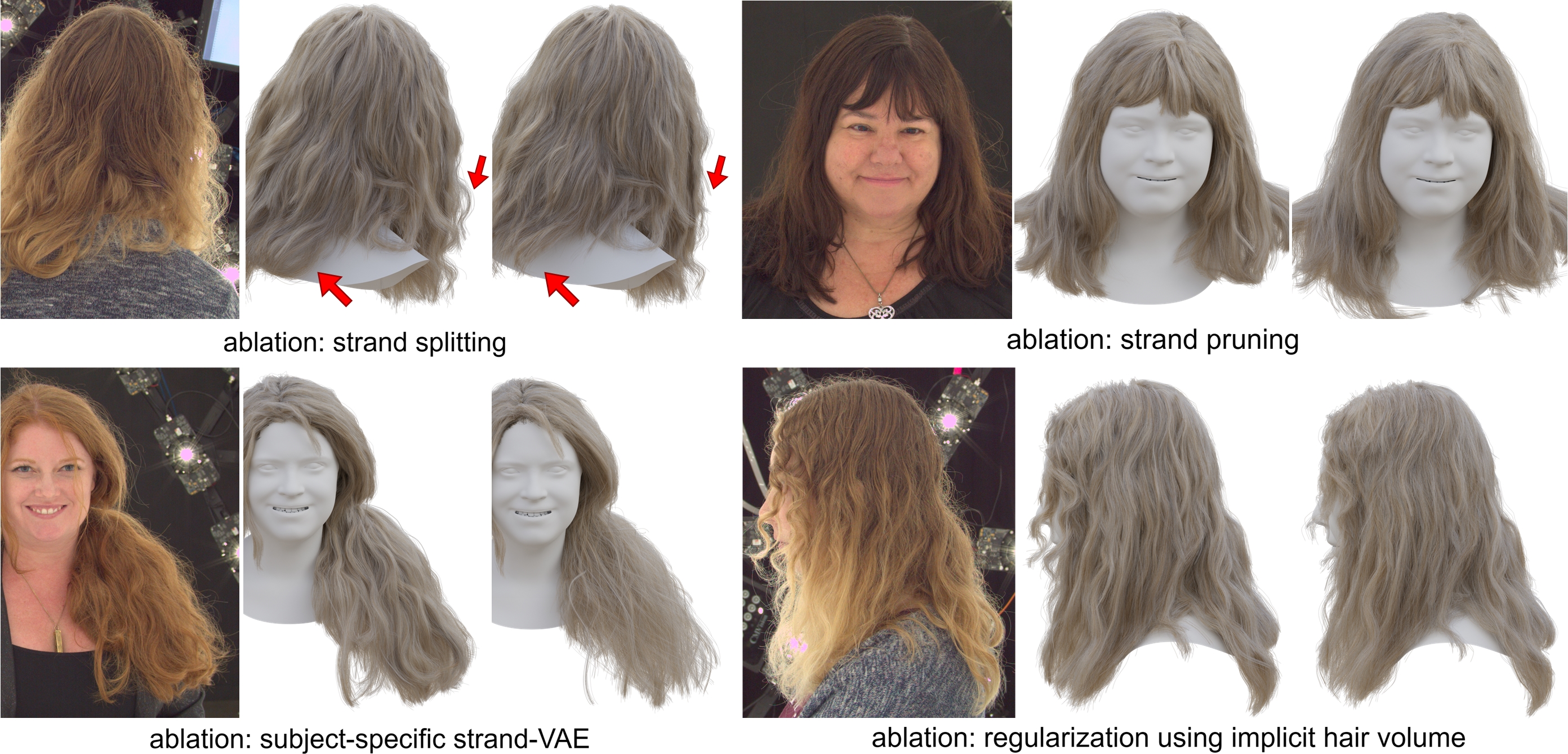}
  \caption{\textbf{Additional ablations for Gaussian-based hair optimization.} Each triplet shows the reference view (left), the result of our full method (middle), and the result of the ablated baseline (right). Top left: the hair without adaptive splitting suffers from worse coverage and wisp structures. Top right: optimization without adaptive pruning leads to excessively long strands. Bottom left: using a pre-trained prior strand-VAE leads to overly smoothed strands due to poor coverage of the synthetic data. Bottom right: regularization with the implicit hair volume helps enhance the hair structure.}
  \label{fig:gs3d_ablation_minor}
\end{figure}

\begin{figure}[h]
  \includegraphics[width=\linewidth]{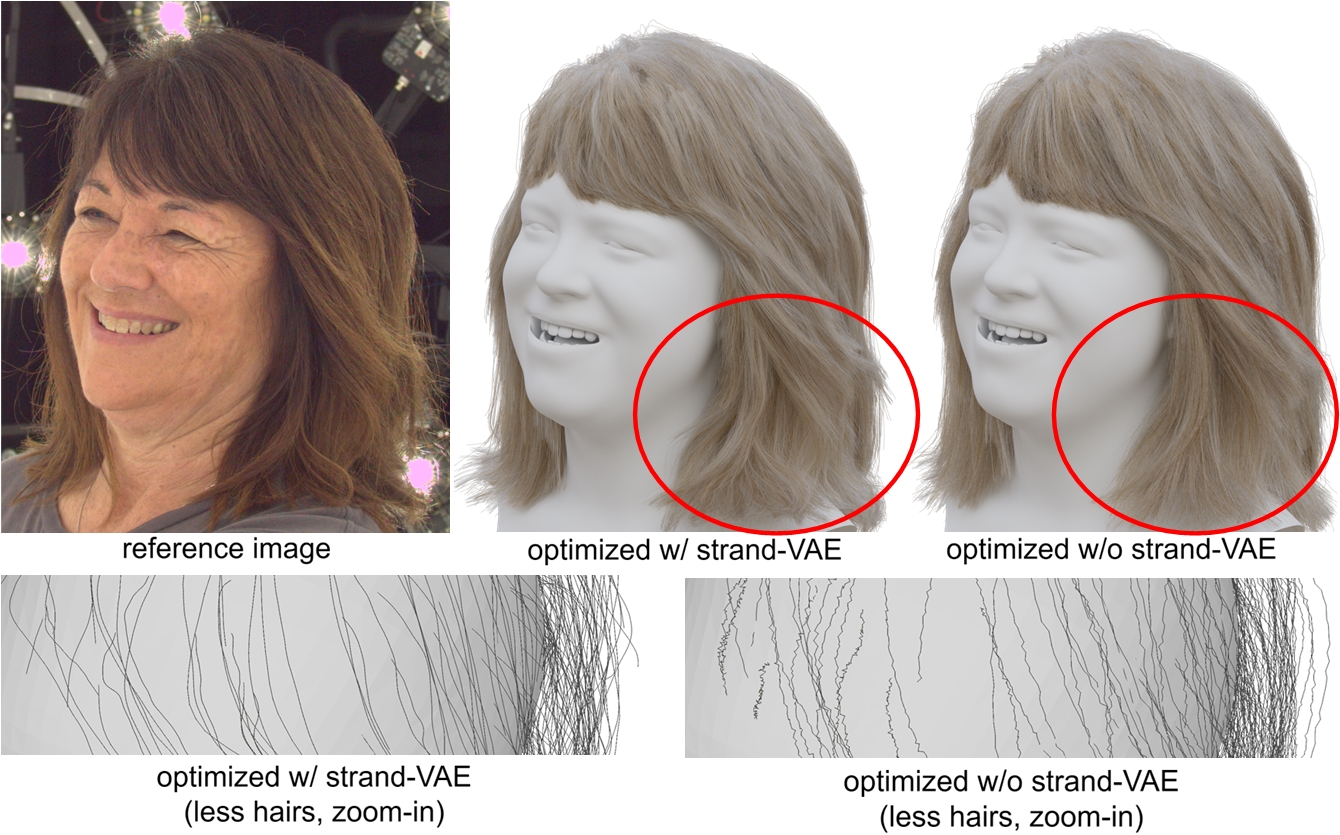}
  \caption{\textbf{Ablation study for the strand latent space.} Optimization within the strand latent space of strand-VAE achieves globally consistent strand deformations that are smoothly regularized (top-middle and bottom-left).
  In contrast, replacing this latent space regularization with a strong smoothing term fails to prevent sharp turns in the strands (bottom-right), even when the overall hair is already overly smoothed (top-right).}
  \label{fig:strand_vae}
\end{figure}

\subsection{Ablation Studies}

In this section we evaluate the critical design choices of our method.
For the implicit hair volume model, our key contributions are: 1) supervising with the full 2D orientation distribution when training the neural orientation network; 2) alpha-blending 3D orientations in histograms when performing volumetric rendering.
To assess these contributions, we train two variants: 1) taking the orientation angles with maximum responses as 2D supervision, instead of the distribution; 2) directly $\alpha$-blending the 3D polar angles, instead of histograms.
As demonstrated in \autoref{fig:nerf_ablation}, supervising only the maximum angles results in locally over-smooth strands because non-maximum orientations are discarded, while blending 3D orientations by directly summing polar angles yields even worse results, as it is mathematically flawed.

In \autoref{fig:dist_3d} we further investigate estimated volume densities and 3D orientations by sampling along an example ray.
The results show that the 3D orientations match different hair layers and lead to correct hair intersections, which is crucial for avoiding over-smoothness for tracing.

We ablate the Gaussian-based optimization stage in \autoref{fig:gs3d_ablation_major}, showing that the optimization leads to improved hair boundaries, more uniform hair density, and more natural strand geometry.
To highlight the effectiveness of Gaussian-based optimization, we process the traced hairs with a low-pass filter and test the optimization on this over-smoothed initialization.
Despite the more challenging input, our optimization successfully recovers most details from the images.
On the contrary, allowing a high DoF as the vanilla 3DGS optimization results in flattened geometry, since tweaking appearance parameters to hallucinate the appearance becomes a shortcut to local optima without genuine geometric details.

We further validate several design choices of the optimization stage in \autoref{fig:gs3d_ablation_minor}.
The first row demonstrates that adaptive strand splitting effectively improves hair coverage and structure, while pruning and cutting are essential to eliminate excess strands.
In the bottom-left, we show that replacing the subject-specific strand-VAE with a generic state-of-the-art model trained on synthetic datasets~\cite{zhou2023groomgen} leads to severe failures due to out-of-distribution strand shapes.
In the bottom-right, we drop both the volume guidance loss $\mathcal{L}_\mathrm{n}$ and the latent regularization loss $\mathcal{L}_\mathrm{\rvl}$ during optimization and observe less structured results.
This suggests that without these terms, which maintains consistency with the implicit hair volume, the Gaussian-based optimization alone does not fully comprehend spatial composition of strands.

Finally, in \autoref{fig:strand_vae}, we justify the necessity of using the strand-VAE during optimization.
In this ablation, we directly optimize the vertex positions without using the strand-VAE and introduce a supplementary smoothing term $\mathcal{L}_\mathrm{s} = | \rve_{i + 1} - \rve_{i} |$.
However, due to the high locality in pixel-wise optimization, the hair strands are severely twisted, even if we enforce a large weight on $\mathcal{L}_\mathrm{s}$ that already leads to global over-smoothness.
In contrast, optimizing the latent vector deforms the strand as a whole, preserving its structural integrity.

\subsection{Applications}
\label{subsec:app}

\begin{figure}[t]
  \includegraphics[width=\linewidth]{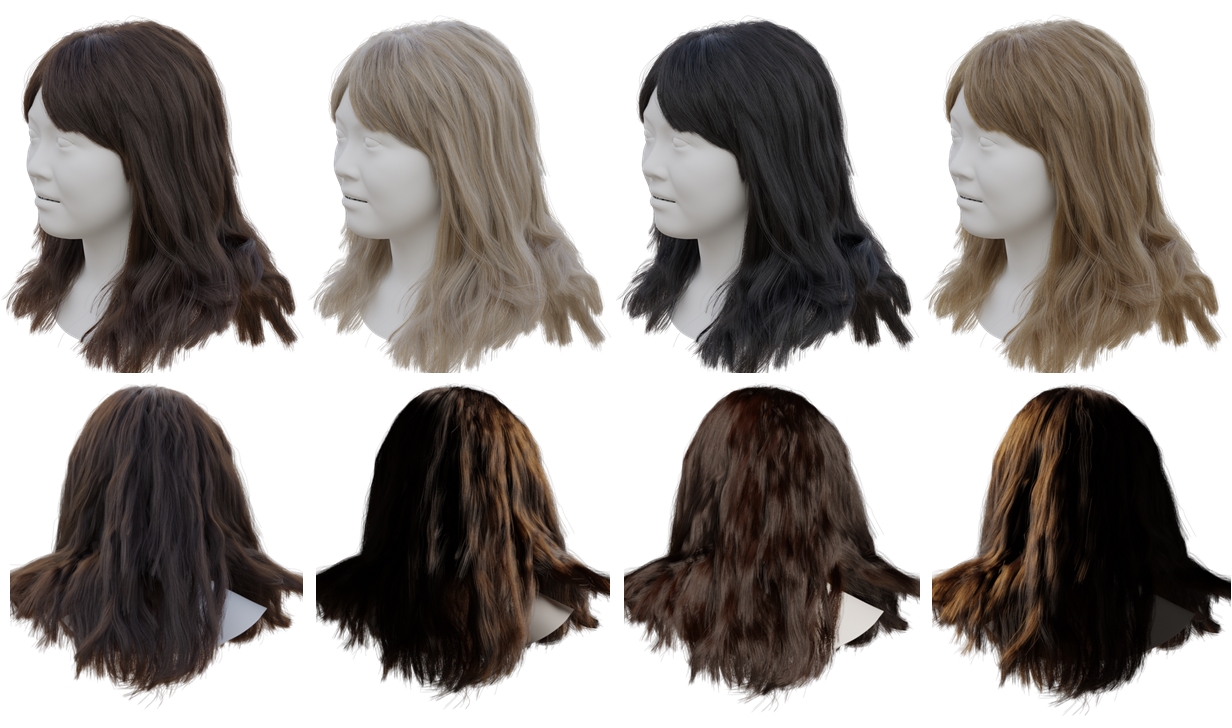}
  \caption{\textbf{Hair re-rendering application.} We render the reconstructed hair geometry using different materials (row 1) and environment lightings (row 2) with a physically-based renderer.}
  \label{fig:rerender}
\end{figure}

\begin{figure}[t]
  \includegraphics[width=\linewidth]{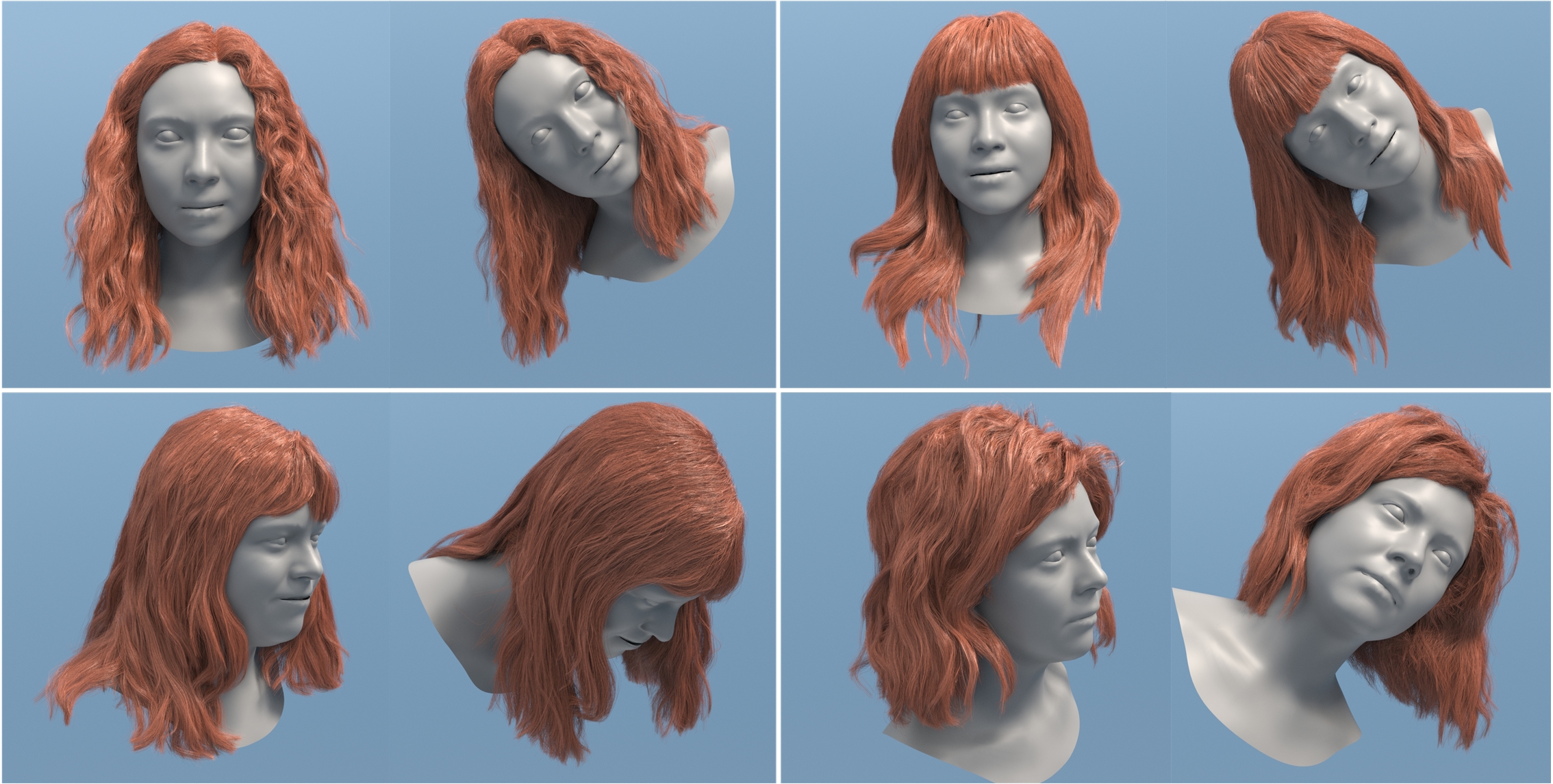}
  \caption{\textbf{Hair simulation application.} In each image pair, we demonstrate the original captured hairs (left), and the hairs deformed with quasi-static simulation at a given head pose. The simulation is performed using the industrial software \textit{Houdini}.}
  \label{fig:simulation}
\end{figure}

\begin{figure}[t]
  \includegraphics[width=\linewidth]{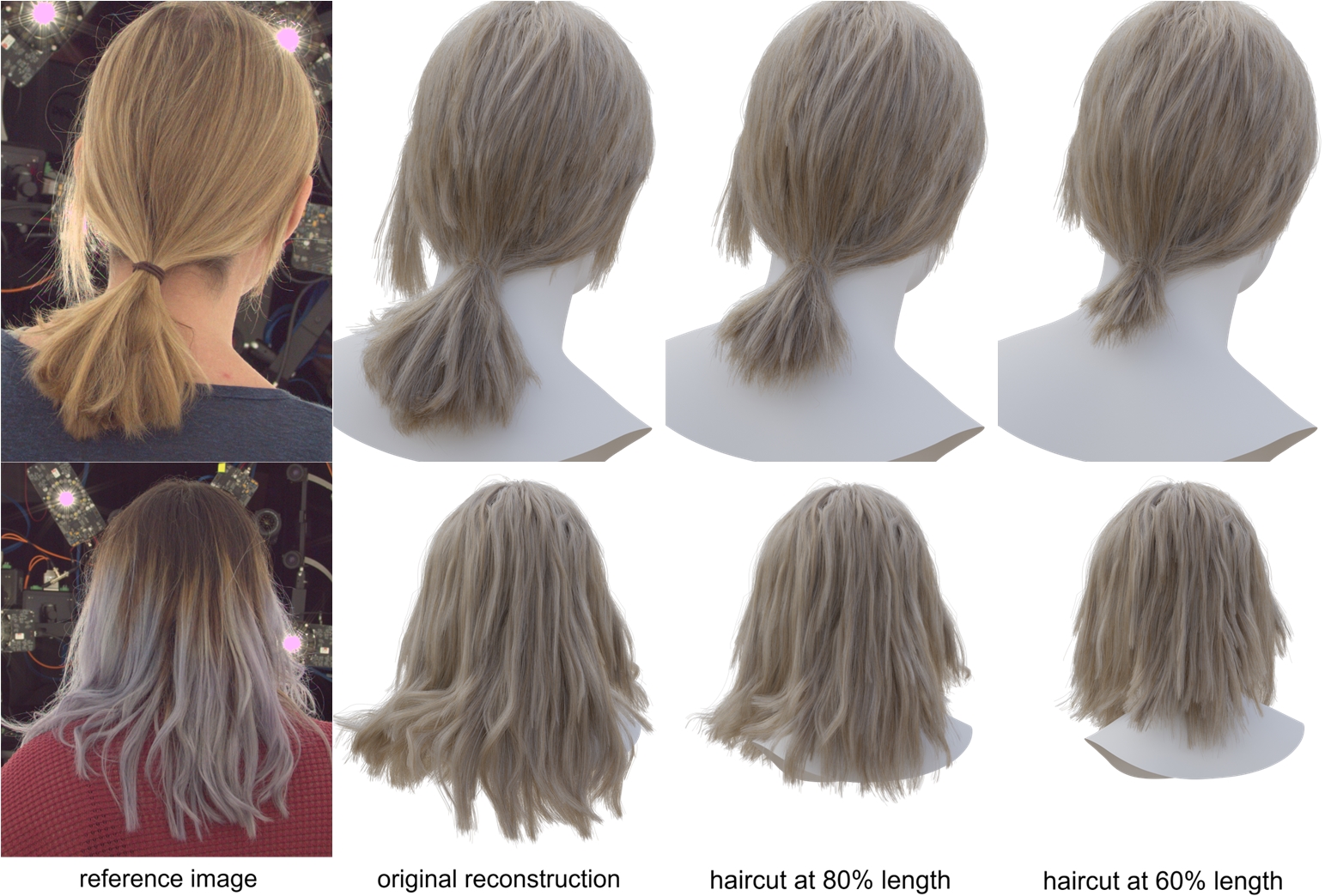}
  \caption{\textbf{Hair editing application.} We perform haircut by keeping 80\% (60\%) of the original vertices for each hair strand.}
  \label{fig:haircut}
\end{figure}

Our method reconstructs explicit hair geometry as a dense set of polyline curves.
Compared to implicit representations of \cite{alexandru22neural,wang23neuwigs}, our reconstruction can be much more easily used in other applications, such as physically-based rendering (\autoref{fig:rerender}), simulation (\autoref{fig:simulation}), and hair editing (\autoref{fig:haircut}).

\subsection{Limitation and Future Work}

\begin{figure}[t]
  \includegraphics[width=\linewidth]{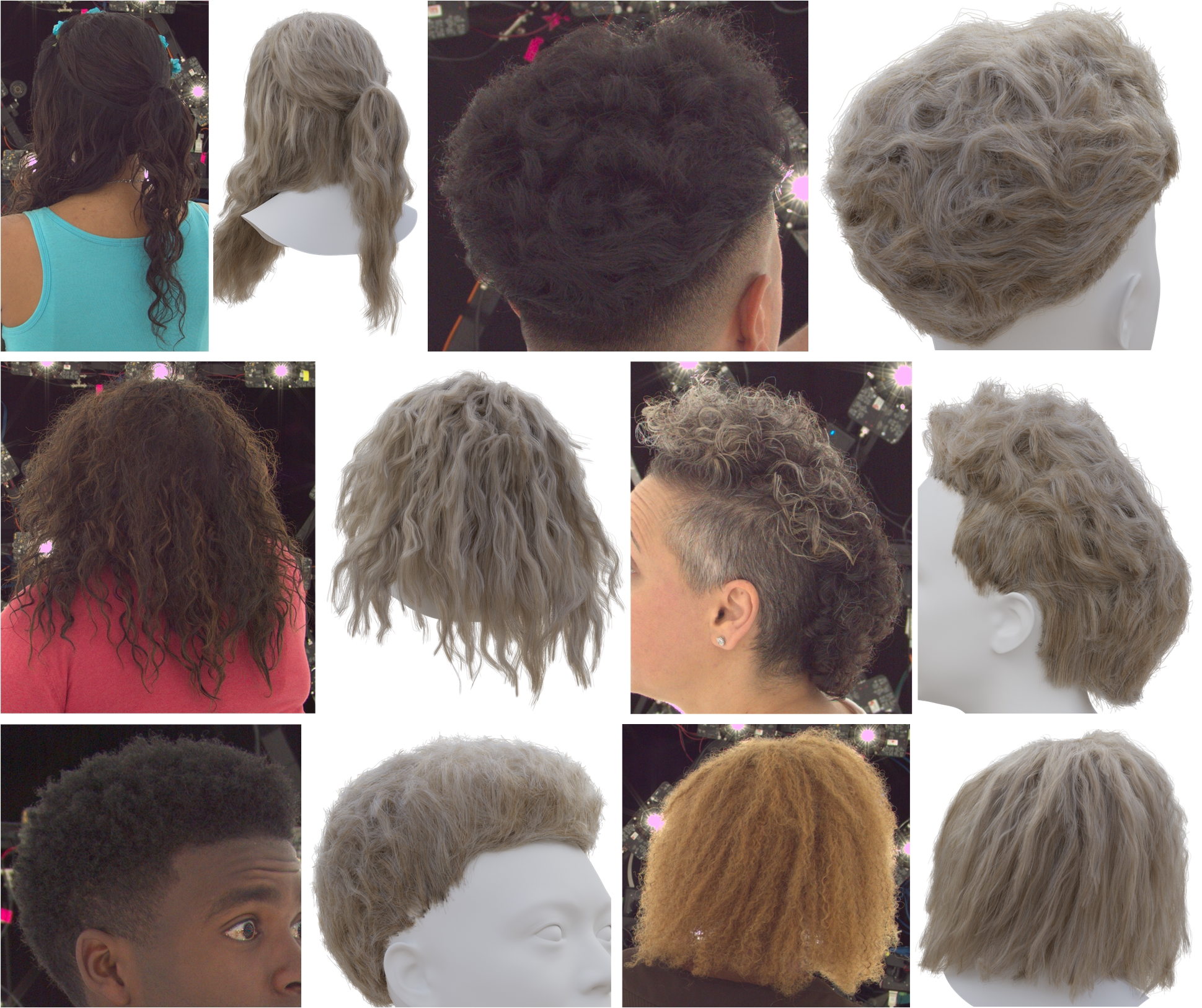}
  \caption{\textbf{Failure cases.} For the extremely complicated hairstyles, our method fails to capture all the small curls.}
  \label{fig:failure}
\end{figure}

While most recent hair capture works heavily rely on prior models derived from synthetic datasets, this paper aims to push the boundaries of prior-free hair capture, targeting diverse hairstyles with rich personal details.
However, we acknowledge the significance of prior information, particularly for handling complex hairstyles.

In \autoref{fig:failure} we present several typical failure cases.
Challenges, arising from dark hair appearances and extremely curly strands, can cause difficulties across all stages: retrieved orientations are noisy, traced strands appear messy, and optimization struggles to effectively enhance the hair quality.
As a result, the final reconstruction may contain hair structures that are inconsistent with the input images.
We believe that integrating prior knowledge with our flexible prior-free capture pipeline represents a promising avenue for future research, as exemplified by MonoHair~\cite{wu24monohair}.

Since our method does not assume any hairstyle priors, it requires reliable segmentation masks to identify the hair regions.
Inaccurate segmentation will lead to hair strands compensating for mislabeled background or body pixels and create misaligned variations that deviate from the input imagery.
Furthermore, as the hair masks are binary, our method cannot fully reflect regional variations in hair density or baldness.
We anticipate that incorporating strand-accurate hair matting will improve the capture of fine details.

\section{Conclusion}

We introduce GroomCap, a novel, prior-free approach for capturing hair geometry from multi-view inputs, effectively bridging the gap between high-fidelity hair modeling and practical application needs.

The first stage of GroomCap involves building a high-resolution implicit hair volume, inspired by neural radiance fields, which incorporates a comprehensive analysis of orientation distributions through volumetric rendering on expanded histograms.
Following the hair volume construction, we trace explicit hairs and utilize 3D Gaussian Splattings for differentiable rendering, facilitating detailed photometric supervision. To refine and regularize the optimization process, hair strands are deformed within a low-dimensional strand latent space, leveraging a subject-specific variational autoencoder. This approach is further enhanced by reduced parameters to prevent appearance hallucinations and adaptive hair splitting/pruning to improve the fidelity of the final hair geometry.

GroomCap has demonstrated its versatility and effectiveness, capturing a diverse range of hairstyles with remarkable quality in both controlled studio and challenging in-the-wild settings.
The success of GroomCap highlights its potential as a transformative tool in various scenarios where high-quality hair is desired.

\begin{acks}
    We thank Vanessa Sklyarova, Keyu Wu, and Youyi Zheng for assistance with comparions, Alessandro Pepe for hair simulation, Di Qiu for hair segmentation, Xu Chen for body fitting, Georgios Kopanas and Chenglei Wu for discussions and proofreading, anonymous reviewers for insightful feedback, and all our capture models.
\end{acks}

% \newpage

\bibliographystyle{ACM-Reference-Format}
\bibliography{ref}

% \clearpage

\appendix
\section{Supplementary Evaluations}
\label{app:evalation}

\begin{figure}[h!]
  \includegraphics[width=\linewidth]{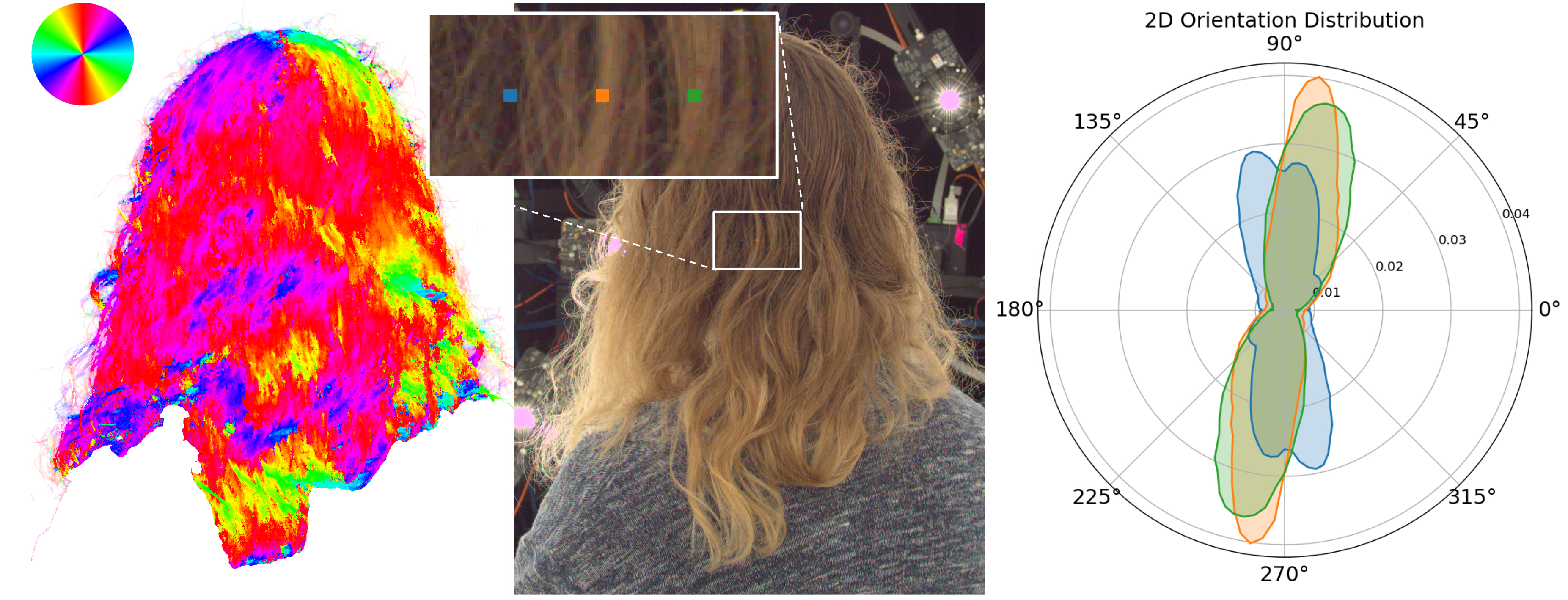}
  \caption{\textbf{Visualization of accumulated 2D orientations.} Left: orientations with highest probabilities after volume rendering. Middle: the reference image with three pixel samples. Right: 2D orientation distributions of the three pixel samples with corresponding colors.}
  \label{fig:dist_2d}
\end{figure}

In this section we provide more evaluations of our method.
In \autoref{fig:dist_2d} we evaluate the implicit hair volume by examining the 2D orientations obtained from volume rendering.
From the right sub-figure, we can see that the peak angles of three distributions are well aligned with hair directions at their sample pixels.
The orange distribution is most concentrated, because of its clean wisp structure, while the blue distribution is most flat due to the blurriness in the image near that pixel.
Overall, our implicit hair volume recovers correct hair structures with the per-pixel orientation distributions faithfully match local hair compositions.

\begin{figure}[h!]
  \includegraphics[width=\linewidth]{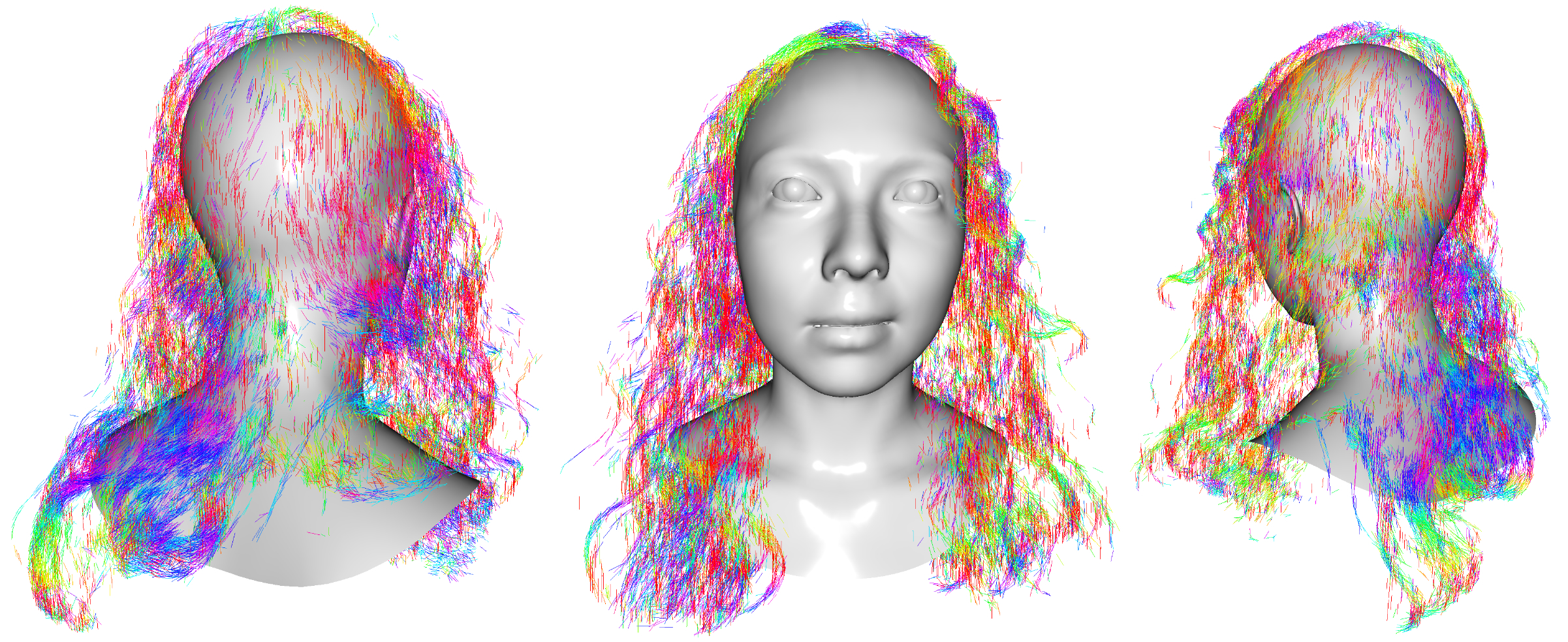}
  \caption{\textbf{Visualization of implicit hair volume} by rendering line segments oriented along the predictions. Reference images in \autoref{fig:inputs}.}
  \label{fig:volume_vis}
\end{figure}

In \autoref{fig:volume_vis}, we further validate our implicit hair volume by drawing line segments with the predicted 3D orientations at point samples with top $5\%$ density values. These sparse line segments already identify the target hairstyle pretty well.

\begin{figure}[H]
  \includegraphics[width=\linewidth]{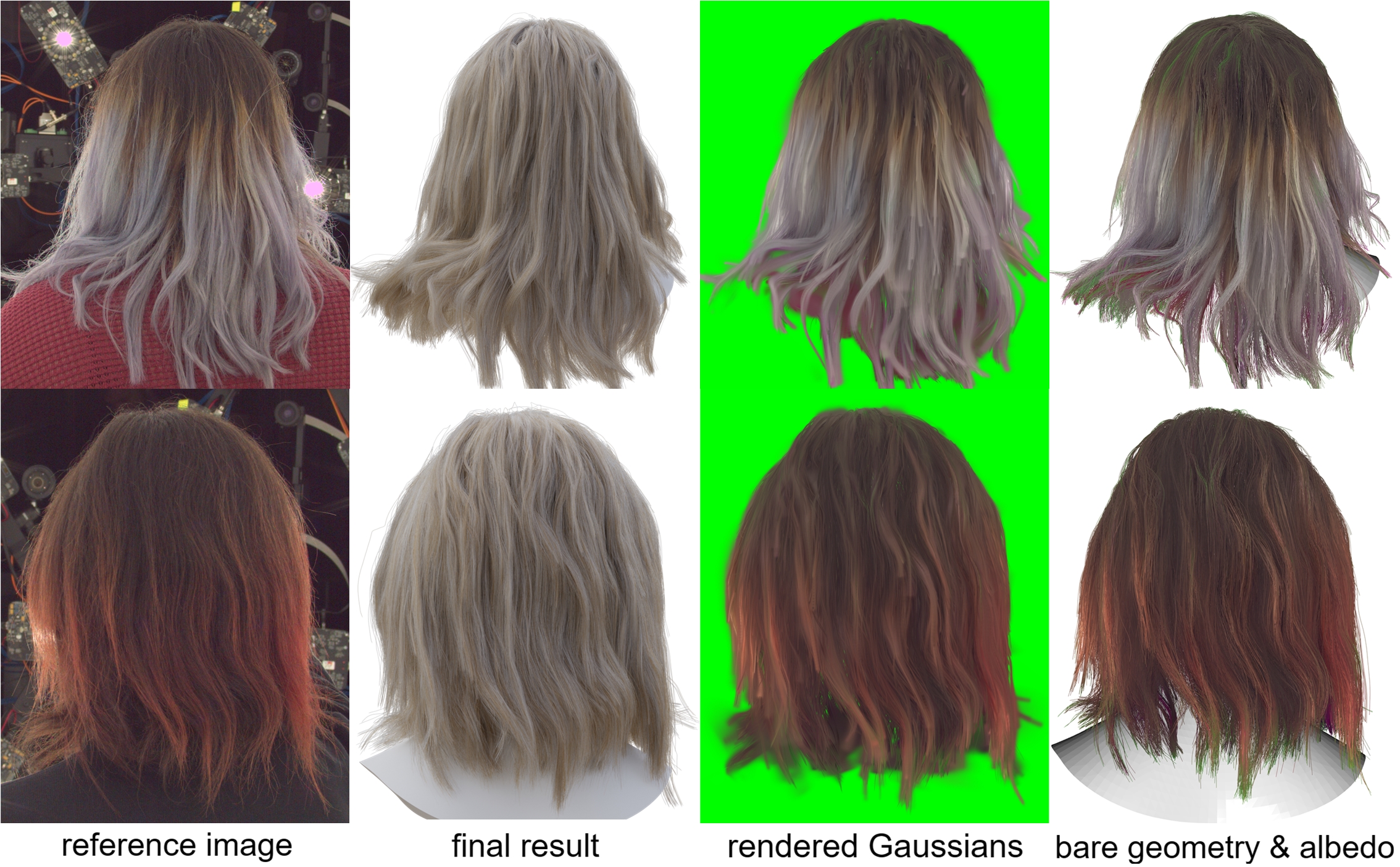}
  \caption{\textbf{Intermediate results of Gaussian-based optimization.} Due to our strong regularization, the rendered hair gaussians do not perfectly reproduce the image. However, the formulation avoids hallucination and leads to improved underlying geometry.}
  \label{fig:gs3d_outputs}
\end{figure}

Finally, in \autoref{fig:gs3d_outputs}, we show the intermediate results of our Gaussian-based hair optimization.
The rendered Gaussians do not perfectly reproduce the input image due to our highly constrained parameters, which helps lead the optimization towards improved geometry.

\section{Hair Segmentation}
\label{app:seg}

We require high-quality hair segmentation masks for all views as the ground truth to train the neural occupancy fields.
While a perfect solution for hair segmentation does not exist, we find that aggregating several off-the-shelf models improves the results.

\begin{figure}[H]
  \includegraphics[width=\linewidth]{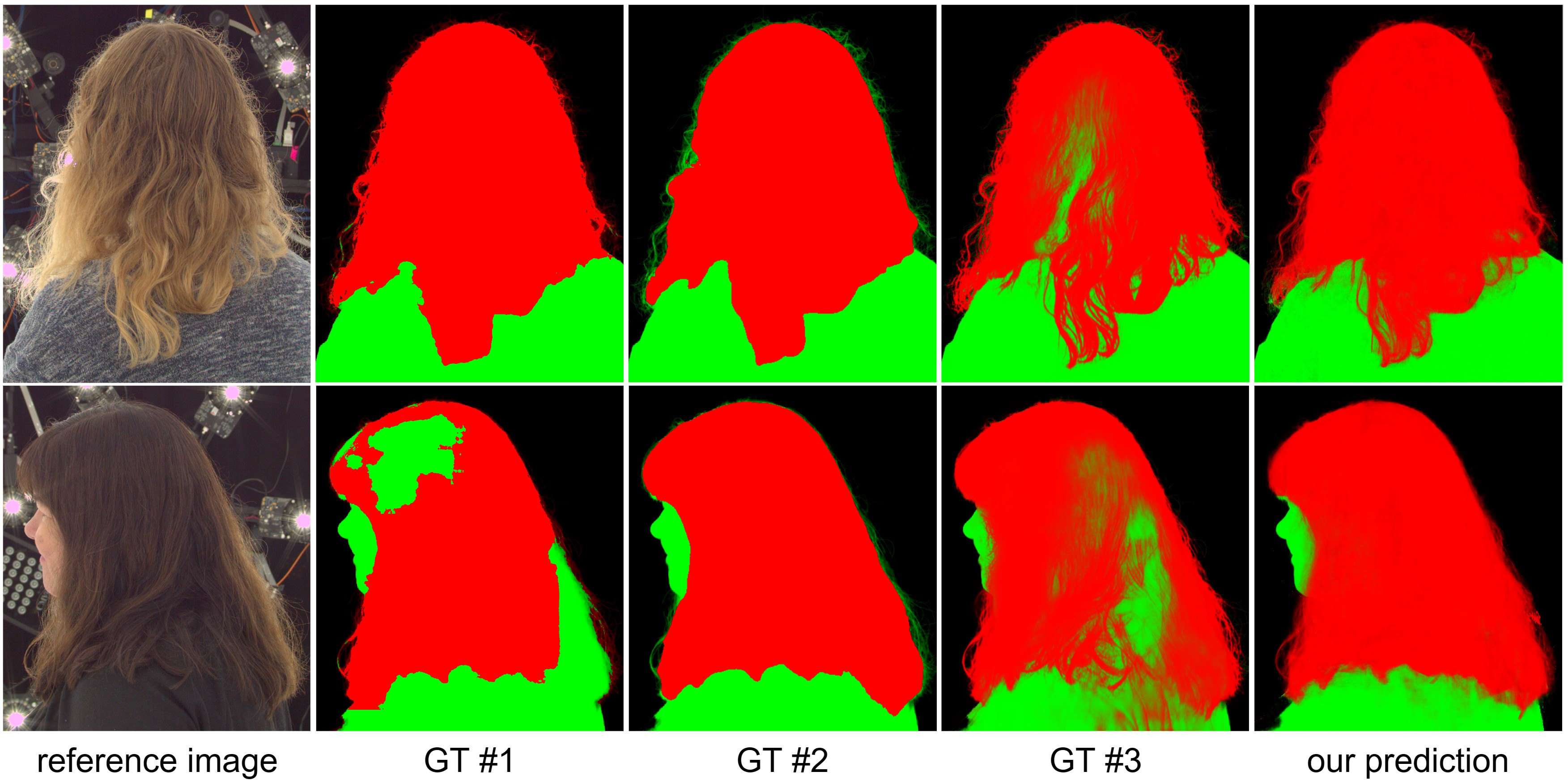}
  \caption{\textbf{Hair segmentation supervision and prediction.} We use pseudo ground truth from multiple sources as supervision to train our neural occupancy model. The model's prediction in turn outperforms the  pseudo ground truth due to the implicit multi-view aggregation.}
  \label{fig:segmentation}
\end{figure}

Specifically, assume we have $m$ segmentation models and denote the per-pixel hair and body likelihood predicted by the $i$-th model as $\bar{\psi}_{h, i}$ and $\bar{\psi}_{b, i}$, we replace the occupancy loss in \autoref{eq:l_occ} with:
\begin{equation}
    \mathcal{L}'_\mathrm{occ} = \min_{i \in \{1, ..., m\}}||\psi_h - \bar{\psi}_{h, i}||^2 + \min_{i \in \{1, ..., m\}}||\psi_b - \bar{\psi}_{b, i}||^2 \mathrm{.}
\end{equation}
Intuitively, we use the supervision that gives the minimal loss, since we experimentally find that the neural occupancy model outperforms all supervisions at the end of training due to its multi-view consistency, as shown in \autoref{fig:segmentation}.
This observation also indicates that our model is robust against incorrect segmentation masks.
In practice, we use $3$ segmentation models, including two in-house models and one public one~\cite{lugaresi2019mediapipe}.

\section{Image Resolution}

\begin{figure}[h]
  \includegraphics[width=\linewidth]{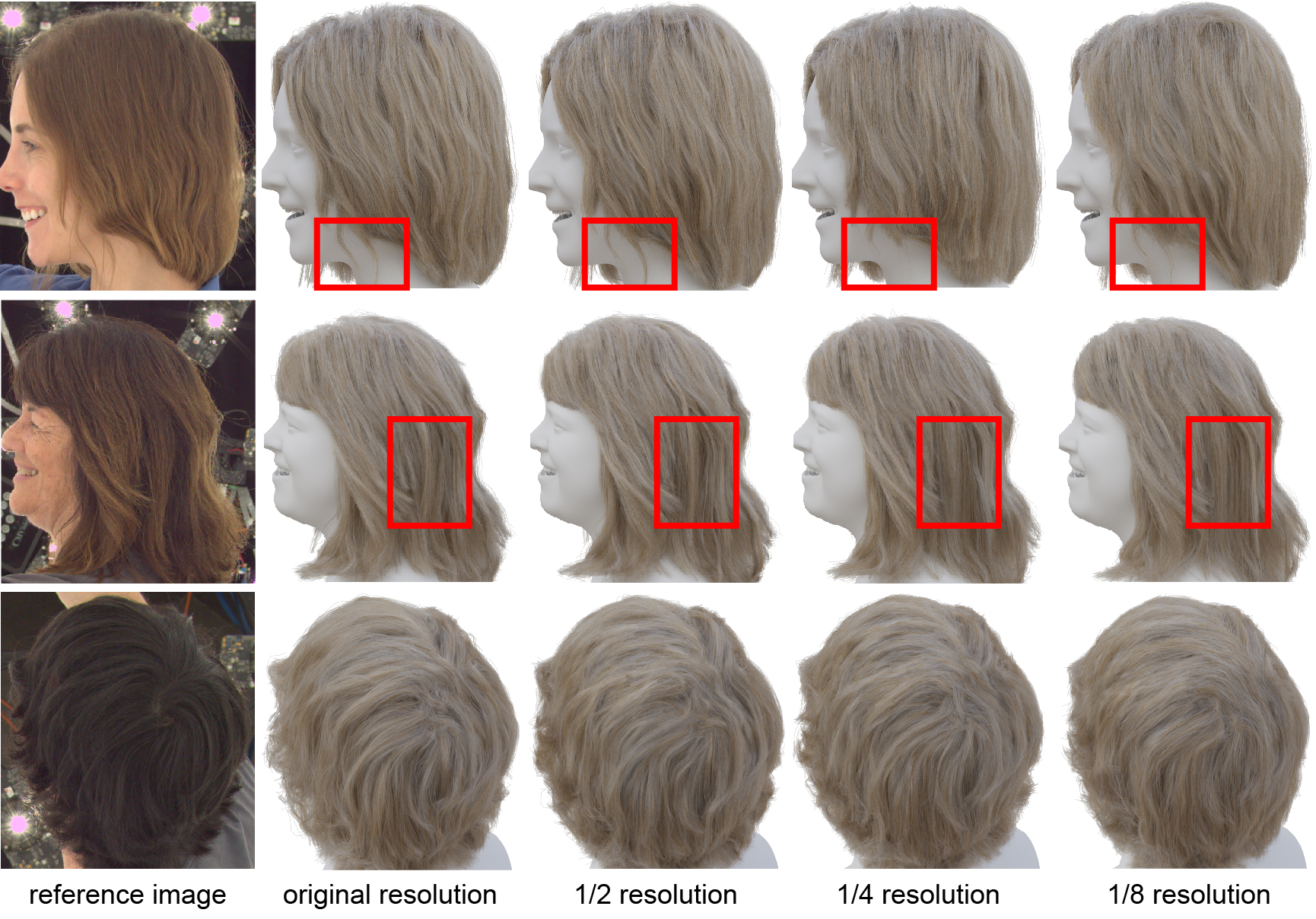}
  \caption{\textbf{Ablation study of different input image resolutions.} With lower-resolution images (three right columns) downsampled from the original captures, while the reconstruction fidelity is reduced due to the loss of fine details in the input, the overall hairstyles are still correctly captured.}
  \label{fig:resolution}
\end{figure}

Our method is robust to image quality.
In \autoref{fig:resolution}, we demonstrate the results reconstructed from lower-resolution images downsampled from the original capture.
While some fine details are inevitably lost due to the reduced input fidelity, the overall structure is still accurately captured, consistent across all resolutions.

\section{Hair Parting Line}
\label{app:partline}

\begin{figure}[h]
  \includegraphics[width=\linewidth]{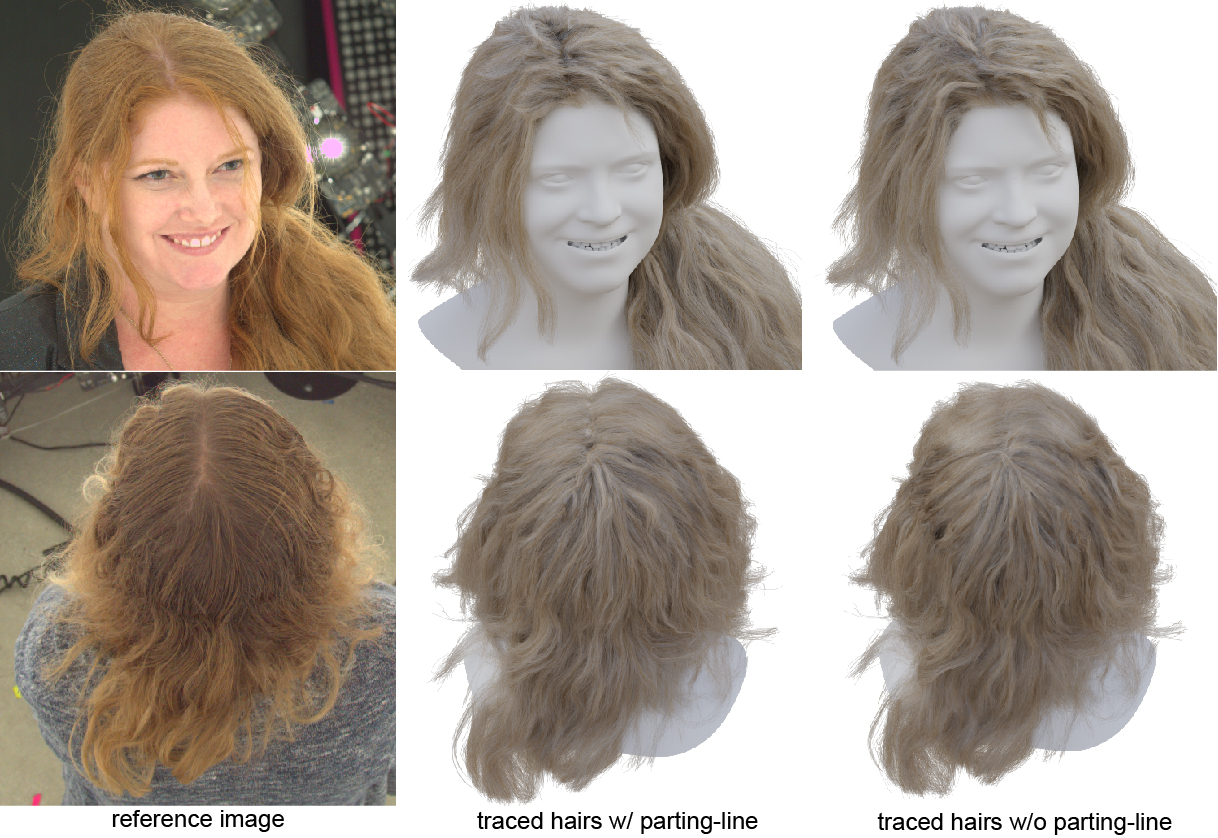}
  \caption{\textbf{Ablation study of parting line annotations in hair tracing.} Without the annotation (right columm), due to the inherent direction ambiguity of orientation estimation, hair strands may grow across the parting line.}
  \label{fig:partline}
\end{figure}

Hair partling line annotations can help our method better mitigate the direction ambiguity.
In \autoref{fig:partline}, we compare the traced hairs with and without parting line annotations.
The results without annotations exhibit blurry parting lines, although structures in other areas remain correct.

\section{Hyperparameters}

As a prior-free method, our pipeline relies on a set of empirically determined constants.
These values are chosen based on our experiments with various capture data, real-world considerations, and common practices in the field.
In the hair tracing stage, we target $125K$ total volume strands, which are sufficient to fill the hair volume, and $25K$ scalp strands, providing enough density to connect the volume strands to the scalp.
During Gaussian-based optimization, we opt for a moderate number of $8$ anchors, which is also divisible by the number of segments.
Increasing the anchors to $15$ yields almost identical results on most captures, while $33$ anchors leads to observable hallucination.
Considering that the memory capacity is around $50K$ strands, the optimization starts from $30K$ hairs to leave room for the adapative control of hair density, allowing for half of the strands to be dynamically created and rearranged.
Notably, our results are not senstive to these particular values, and we use the same set of hyperparameters for every diverse hairstyle we demonstrate in the paper.

% Should be 15 pages.

\end{document}